\documentclass[
twocolumn,
preprintnumbers,
nofootinbib, 
showpacs,
prd,
aps,
]{revtex4-2}
\pdfoutput=1
\usepackage{graphicx}
\usepackage{amsmath, amssymb}
\usepackage{slashed}
\usepackage{xcolor}
\usepackage{array,arydshln}
\usepackage[pdftex,
  colorlinks=true,
  citecolor=green!60!blue]{hyperref}

\usepackage[italicdiff]{physics}

\begin{document}

\preprint{KANAZAWA-21-02}

\title{
Axion mass in antiferromagnetic insulators
}

\author{Koji Ishiwata}

\affiliation{Institute for Theoretical Physics, Kanazawa University,
  Kanazawa 920-1192, Japan}

\date{\today}

\begin{abstract}
  We calculate axion potential in antiferromagnetic insulators from
  path integral.  It is derived from the effective potential for the order
  parameter of the antiferromagnetic phase in insulators.  {\it
    Static} and {\it dynamical} axions are defined consistently from
  the  potential. Consequently antiferromagnetic/paramagnetic and
  topological/normal orders are classified. The dynamical axion is
  predicted in all phases and its mass turns out to be various values
  up to eV scale.

\end{abstract}

\maketitle

\section{introduction}
\label{sec:intro}

Axion is a hypothetical elementary particle that is considered to
solve the strong CP problem in QCD~\cite{Peccei:1977hh}. It is
originally massless due to shift symmetry of the Lagrangian, and it
becomes massive after the QCD transition via a nonperturbative
effect. Other than the QCD axion, axion particles are known to emerge
from superstring theory.  Motivated partly by that, axion-like
particles (ALPs) are studied in phenomenological, astrophysical, and
cosmological contexts. The mass of ALPs is usually considered as a
free parameter, leading to more rich phenomenology compared to the QCD
axion (see, {\it e.g.}, Ref.\,\cite{Zyla:2020zbs} for review).

Besides in the field of particle physics, axion is discussed in
condensed matter physics. For example, the magnetoelectric response
due to $\vb*{E}\cdot \vb*{B}$ coupling ($\vb*{E}$ and $\vb*{B}$ are
electric and magnetic fields, respectively) is already known, and the
$\vb*{E}\cdot \vb*{B}$ coupling {\it is} the interaction term of axion
with the gauge field. Therefore, such magnetoelectric response may be
able to understood in terms of ``axion.'' There have been several
studies in that context so far.  Different from axion in elementary
particle physics, there are two types of ``axions'' in condensed
matter physics: {\it static} axion and {\it dynamical}
axion. (Hereafter we use ``axion'' as an axion in condensed matter
physics unless otherwise mentioned.) In terms of Lagrangian, they are
written as
\begin{align}
  {\cal L}_\theta =
    \frac{\alpha}{\pi}(\theta_0+a)\vb*{E}\vdot\vb*{B}\,,
\end{align}
where $\alpha=e^2/(4\pi)$ is the fine-structure constant and we have
denoted $\theta_0$ as the static axion (constant) and $a$ as the
dynamical axion.  The former is pointed out by
Refs.\,\cite{Qi:2008ew,Qi:2008pi,Hehl:2007ut,Dzyaloshinskii,Essin:2009,Wu:2016}
as the $\vb*{E}\cdot \vb*{B}$ coupling. Nonzero $\theta_0$ breaks both
time-reversal ${\cal T}$ and parity ${\cal P}$. However, when
$\theta_0=\pi$ those symmetries are intact. This nontrivial state is
known as an topological state.  Ref.\,\cite{Li:2010}, on the other
hand, studied the latter to predict that the dynamical axion causes
the total reflection of light irradiated to the topological insulators
under an antiferromagnetic (AFM) order. They solved equations of
motion of photon and the dynamical axion that interact with each other
via the $\vb*{E}\vdot \vb*{B}$ interaction term and found an energy
gap emerged in the topological insulators.  In the similar context,
Ref.\,\cite{Ooguri:2011aa} shows that an electric field is expected to
be converted to a magnetic field due to the dynamical axion when the
electric field exceeds certain critical value.  (See also
Refs.\,\cite{Wang:2015hhf,Imaeda:2018,Taguchi:2018hge,Varnaga:2018,Sekine:2020ixs}.)
Such recent developments in both particle physics and condensed matter
physics have motivated {\it particle} axion and ALPs search by using
axions in
insulators~\cite{Marsh:2018dlj,Chigusa:2020gfs,Schutte-Engel:2021bqm,Chigusa:2021mci}.

In this paper we calculate the effective potential for the order
parameter of the AFM phase from the partition function of the system
defined by the path integral. Using the effective potential, axion
potential is derived.  It is found that the effective potential and
axion potential are intuitive and useful in order to distinguish the
AFM/paramagnetic (PM) order and the topological/normal phase depending on
the model parameters.  We clarify the static and dynamical axions in
the AFM insulators and how they are related to the AFM order and
the topological state.  Consequently, the mass of dynamical axion is
determined. While it is estimated to be $\order{\rm meV}$ in
Ref.\,\cite{Li:2010}, the axion mass turns out to be larger up to
$\order{{\rm eV}}$ or a much suppressed value depending on the
property of insulators.

This paper is organized as follows. In Sec.\,\ref{sec:V_phi}, the
effective potential for order parameter of the AFM phase is
derived. Using the result, the AFM/PM phases are determined. The effective
potential is interpreted in terms of axion field and axion potential
is derived in Sec.\,\ref{sec:V_theta}. Sec.\,\ref{sec:conclusions}
contains conclusions and discussion.

\section{potential of antiferromagnetic order parameter }
\label{sec:V_phi}

The topological state in insulators is consider to be realized by a (quasi)
gapless state via the spin-orbit coupling. For modeling such a state,
a minimum setup is to consider 4-by-4 matrix, {\it i.e.}, two energy
states (or orbits or sublattices) with up and down spins.  Such a
model is often described by the gamma matrices (even in lower or
higher dimension space~\cite{Qi:2008ew}). In the momentum space, it is
given by
\begin{align}
  H(\vb*{k})=\epsilon_0(\vb*{k}) {\bf 1}_{4\times 4}
  +\sum_{a=1}^{5}d^a(\vb*{k}) \Gamma^a\,,
  \label{eq:Hk}
\end{align}
where $\vb*{k}=(k_x,\,k_y,\,k_z)$ is the wavenumber and $\Gamma^a$ are
the gamma matrices satisfying
\begin{align}
  \{\Gamma^a,\Gamma^b\}=&\,2\delta^{ab}{\bf 1}_{4\times 4}\,,
  \label{eq:Gamma_formula_1} \\
  {\rm tr}(\Gamma^a\Gamma^b)=4\delta^{ab}\,,
  &~~~{\rm tr}(\Gamma^a) =0\,.
  \label{eq:Gamma_formula_3}
\end{align}
The representation of the gamma matrices depends on the basis. In
Refs.\,\cite{Li:2010,Zhang:2009zzf}, for example, the basis of the
antibonding and bonding are chosen, {\it i.e.}, ($\ket{
  P1_z^+,\sigma}$, $\ket{P2_z^-,\sigma}$)
($\sigma=\uparrow,\,\downarrow$):
\begin{align}
  \Gamma^a=
  (\sigma^x\otimes
s^x,\sigma^x\otimes s^y,\sigma^y\otimes{\bf 1},\sigma^z\otimes {\bf
  1},\sigma^x\otimes s^z)\,,
\label{eq:Gammas}
\end{align}
where $\sigma^{j}$ and $s^{j}$ ($j=x,y,z$) are the Pauli matrices.
The coefficients $d^a$, on the other hand, are given by theoretical
model and/or the first-principles computation.  Among models, the
so-called Dirac model is a low energy effective model to describe the
topological insulators expanding a Hamiltonian around a Dirac
point. In the Dirac model, $d^a$ are given as
\begin{align}
  (d^1,\,d^2,\,d^3,\,d^4,\,d^5)
  =(Ak_x,\,Ak_y,\,Ak_z,\,m+Bk^2,m_5)\,,
  \label{eq:DiracModel}
\end{align}
where $k^2=|\vb*{k}|^2$ and the gamma matrices are given in
Eq.\,\eqref{eq:Gammas}.  $\epsilon_0$, $A$, $B$, $m$, and $m_5$ have a
mass dimension of one. ($\epsilon_0$ is not important in our later
discussion, which is explained in Appendix \ref{sec:EL}.) The Dirac
model can be derived from a certain model, such as an extended
Fu-Kane-Mele-Hubbard model on a diamond lattice at half-filling,
discussed in Refs.\,\cite{Sekine:2014xva,Chigusa:2021mci}, or the
first-principles calculation of the layered, stoichiometric crystals
Sb$_2$Te$_3$, Sb$_2$Se$_3$, Bi$_2$Te$_3$ and
Bi$_2$Se$_3$~\cite{Zhang:2009zzf}.  In general, the coefficient $A$
corresponds to the spin-orbit coupling.\footnote{To be precise, the
spin-orbit interaction itself is local and it is constant in momentum
space. The momentum dependence comes from nonlocal interaction
coupled to spin-orbit coupling. } If it is zero, then $\theta$ is
exactly zero and no topological phase appears, which will be easily
understood from Eq.\,\eqref{eq:theta_phi}. The coefficient $m$ and $B$
describe the energy gap. They originate in, for instance, the
spin-orbit coupling~\cite{Zhang:2009zzf}, or hopping strength
anisotropy due to the lattice distortion in the diamond
lattice~\cite{Sekine:2014xva}.  $m$ (and $B$) effectively determines
whether the insulator is topological or normal.  Namely, $m/B<0$
($m/B>0$) corresponds to the topological (normal) state for $B\neq
0$. When $B=0$, negative (positive) $m$ gives the topological (normal)
state.  Finally $m_5$ is an order parameter of the AFM phase and plays
an important role for describing axion in insulators, which will be
discussed in detail soon.

While in the numerical study we adopt the Dirac model, the following
analytical calculation is applied for generic model described by
Eq.\,\eqref{eq:Hk}. Besides, in Appendix~\ref{sec:results_in_Limodel}
we give another numerical results in the effective model for 3D
topological insulators given in
Eq.\,\eqref{eq:LiModel}~\cite{Li:2010}.

The $m_5\Gamma^5$ term appears due to the Hubbard-Stratonovich
transformation or the mean field approximation of the Hubbard
interaction term, which causes the AFM order written in a sublattice
basis ($\ket*{{\rm A},\sigma}$, $\ket*{{\rm B},\sigma}$),
\begin{align}
  {\cal H}_{{\rm int}} 
  =
  \frac{UV}{N}\int d^3x~
  (n_{{\rm A}\uparrow}n_{{\rm A}\downarrow}
  +n_{{\rm B}\uparrow}n_{{\rm B}\downarrow})\,,
  \label{eq:H_int}
\end{align}
where $n_{{\rm A}\sigma}=\psi^\dagger _{{\rm A}\sigma}\psi_{{\rm
    A}\sigma}$ and $n_{{\rm B}\sigma}=\psi^\dagger_{{\rm
    B}\sigma}\psi_{{\rm B}\sigma}$ are number density of electron with
spin $\sigma =\uparrow,\,\downarrow$ at sublattice A and B,
respectively. $U$ is a parameter with mass dimension one, $V$ is the
volume of the material and $N$ is the number of sublattice $A$ ($B$)
in the material. This interaction term involves four electrons and it
is difficult to analyze. However, it can be rewritten by the
Hubbard-Stratonovich transformation with introducing a scalar field
$\phi$ in the path integral. (See Appendix~\ref{sec:m5term} for
details.) As a result, the interaction term becomes
\begin{align}
  &\exp\left[-i \int dt{\cal H}_{{\rm int}}  \right]
  \nonumber \\
 &=
  \int {\cal D}\phi
  \exp\left\{-i
  \int d^4x \left[ M^2 \phi^2 +
    \phi
    (n_{{\rm A}-}-n_{{\rm B}-})
     \right]+\cdots
  \right\}\,,
\end{align}
where $n_{{\rm I}-}=n_{{\rm I}\uparrow}-n_{{\rm
    I}\downarrow}$ (${\rm I}=$A, B), and
\begin{align}
  M^2
  =\int \frac{d^3k}{(2\pi)^3}\frac{2}{U}
  \,.
  \label{eq:M^2}
\end{align}
Here the momentum integral is executed in the first Brillouin zone
unless otherwise mentioned, which is applied to other momentum
integral in the later discussion.  The term proportional to $n_{{\rm
    A}-}-n_{{\rm B}-}$ gives $m_5\Gamma^{5\prime}$ term in the
sublattice basis. (See Appendix~\ref{sec:theta_chiral} for the
explicit expression of $\Gamma^{a\prime}$.) Here we have renamed $m_5$ as
\begin{align}
  \phi\equiv m_5\,.
\end{align}
When $m_5=0$ the Hamiltonian is ${\cal T}$
and ${\cal P}$ conserving while a finite value of $m_5$ breaks both.
In addition, it should be noticed that the mass term for $\phi$
emerges, which must be included in the effective Lagrangian for
$\phi$.

For later convenience we separate the Hamiltonian as
\begin{align}
  H(\vb*{k})=H_0(\vb*{k}) + \delta H(\vb*{k})\,,
\end{align}
where
\begin{align}
  H_0(\vb*{k})&=\epsilon_0(\vb*{k}) {\bf 1}_{4\times 4}
  +\sum_{a=1}^{4}d^a_0(\vb*{k}) \Gamma^a\,,
  \label{eq:H0}\\
  \delta H(\vb*{k})&=m_5\Gamma^5\,.
  \label{eq:deltaH}
\end{align}
Here $d_0^a$ is defined by
\begin{align}
  d_0^a=(d^1,\,d^2,\,d^3,\,d^4,\,0)\,.
\end{align}
From the Hamiltonian, the partition function is defined by functional
integral of four-component electron field $\psi$ and $\phi$,
\begin{align}
    Z=\int {\cal D}\psi {\cal D}\psi^\dagger {\cal D}\phi~
  e^{iS+iS_{\phi}^{\rm mass}}\,,
  \label{eq:Z}
\end{align}
where 
\begin{align}
  &S=\int \dd[4]{x} \psi^\dagger(x)\left[i\partial_t-H\right] \psi(x)\,,
  \\
  &S_{\phi}^{\rm mass}=-\int \dd[4]{x} M^2\phi^2\,.
  \label{eq:S_mass}
\end{align}
Here $H$ is the Hamiltonian in the coordinate space.  Now we integrate
out $\psi$ in Eq.\,\eqref{eq:Z} to derive the effective Lagrangian for
$\phi$. It is straightforward to get
\begin{align}
  Z&= \int {\cal D}\phi~ e^{iS_{\phi}^{\rm mass}}\,{\rm det}
  \left[-i\partial_t +H\right]
    \nonumber \\
  &=\int {\cal D}\phi~
              {\rm exp}\left\{iS_{\phi}^{\rm mass}+
              {\rm Tr} \log \left[-i\partial_t +H\right] \right\}\,,
\end{align}
Here ``Tr'' includes that trace over matrices and spacetime
integral. To further rewrite the above expression, let us define
Green's function as
\begin{align}
  G_0^{-1}\equiv i\partial_t -H_0\,,
\end{align}
where $H_0$ is $H_0(\vb*{k})$ written in the coordinate space. Then,
separating $H$ as $H=H_0+\delta H$ and using
\begin{align}
  \log\left[-i\partial_t +H\right]
  &=\log(-G_0^{-1})-\sum_{n=1}^\infty \frac{1}{n}(G_0 \delta H)^n\,,
\end{align}
we get
\begin{align}
  Z&=\int {\cal D}\phi\, {\rm exp}\Bigl\{iS_{\phi}^{\rm mass}+
  {\rm Tr}\left[ \log(-G^{-1}_0) \right]
  \nonumber \\  &~~~~~~
  -{\rm Tr}\Bigl[\sum_{n=1}^\infty \frac{1}{n}(G_0 \delta H)^n\Bigr]
  \Bigr\}\,.
  \label{eq:Z_in_expand}
\end{align}

\begin{figure*}[t]
  \begin{center}
    \includegraphics[scale=0.38]{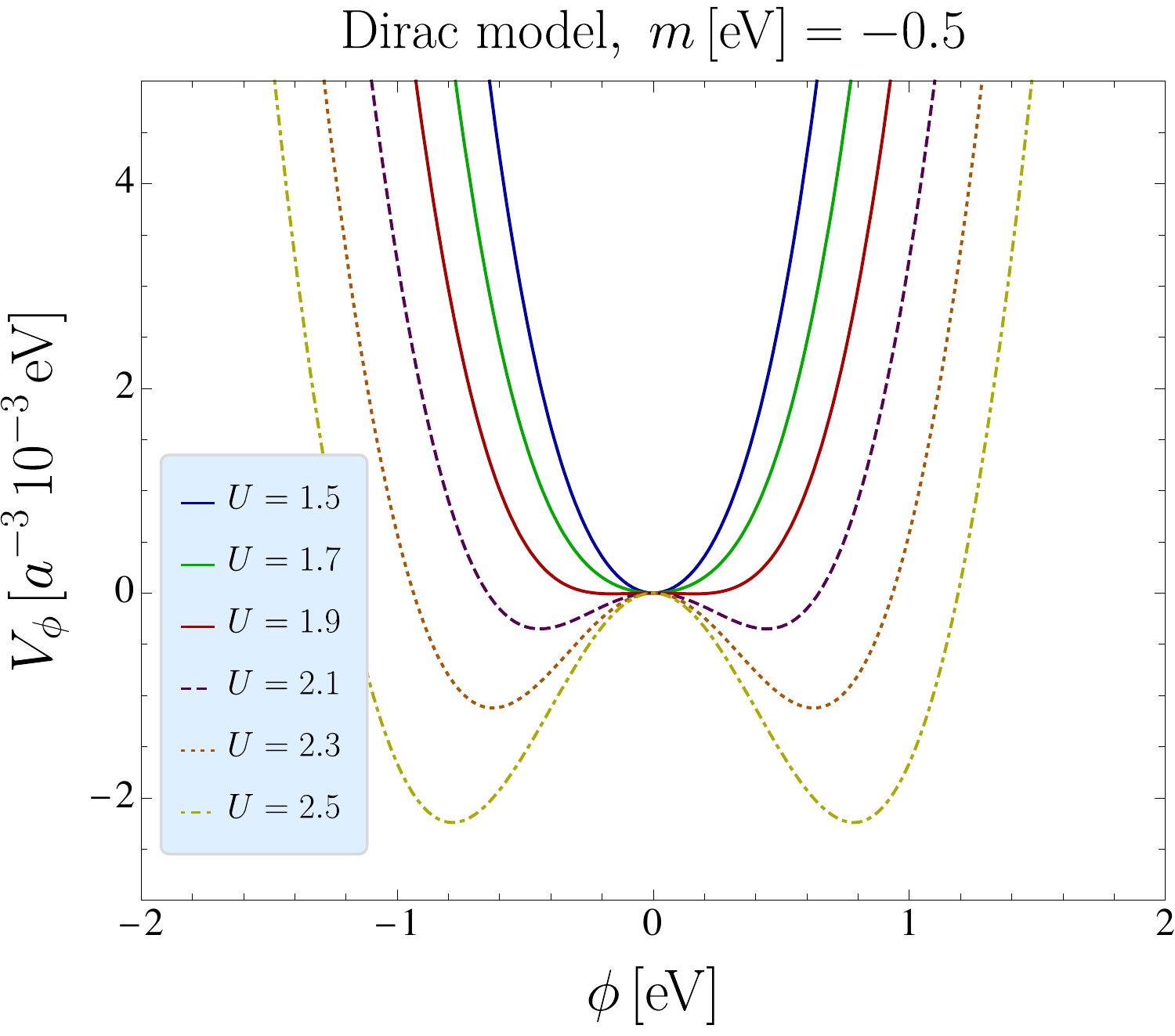}
    \includegraphics[scale=0.38]{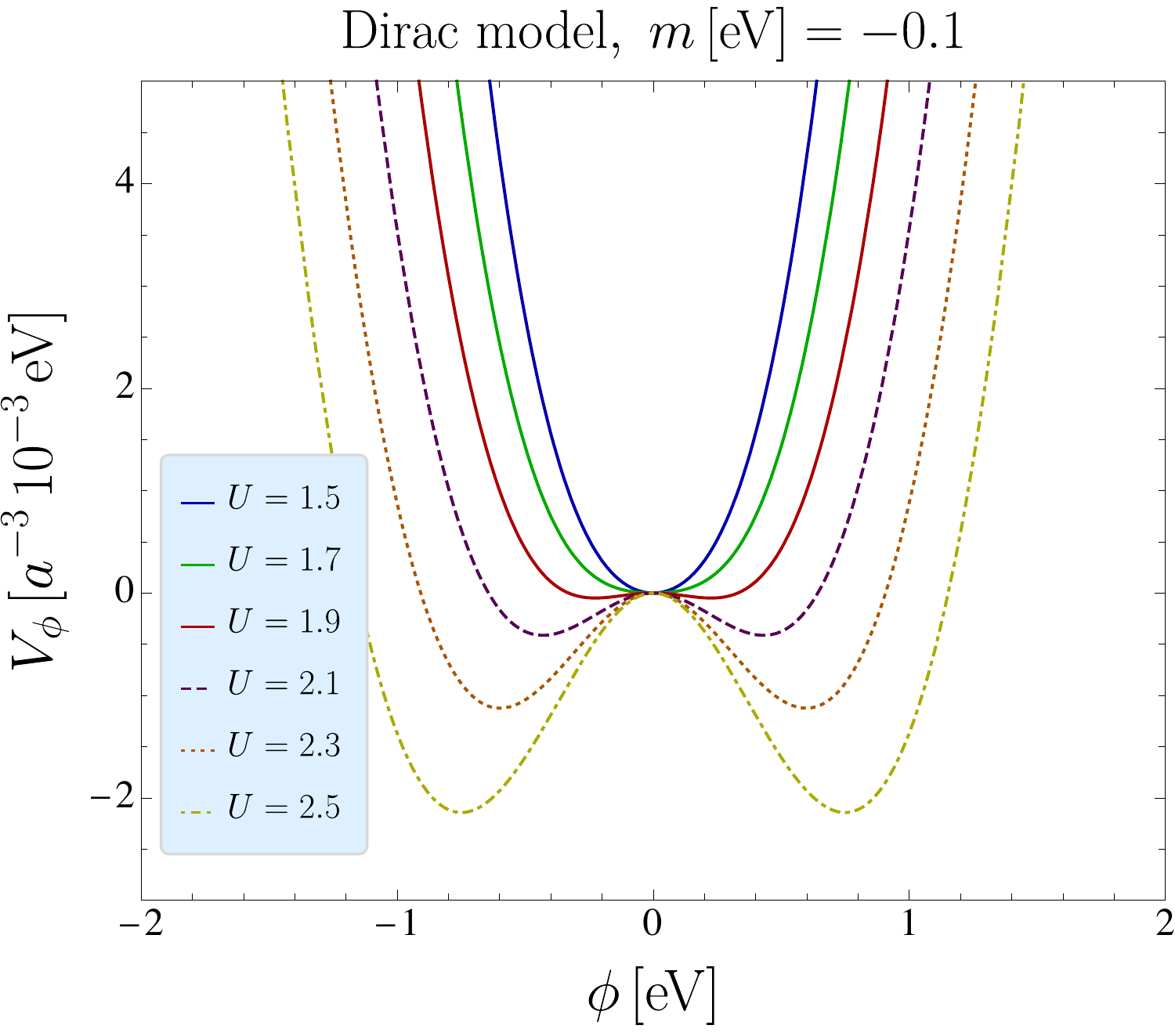}
    \includegraphics[scale=0.38]{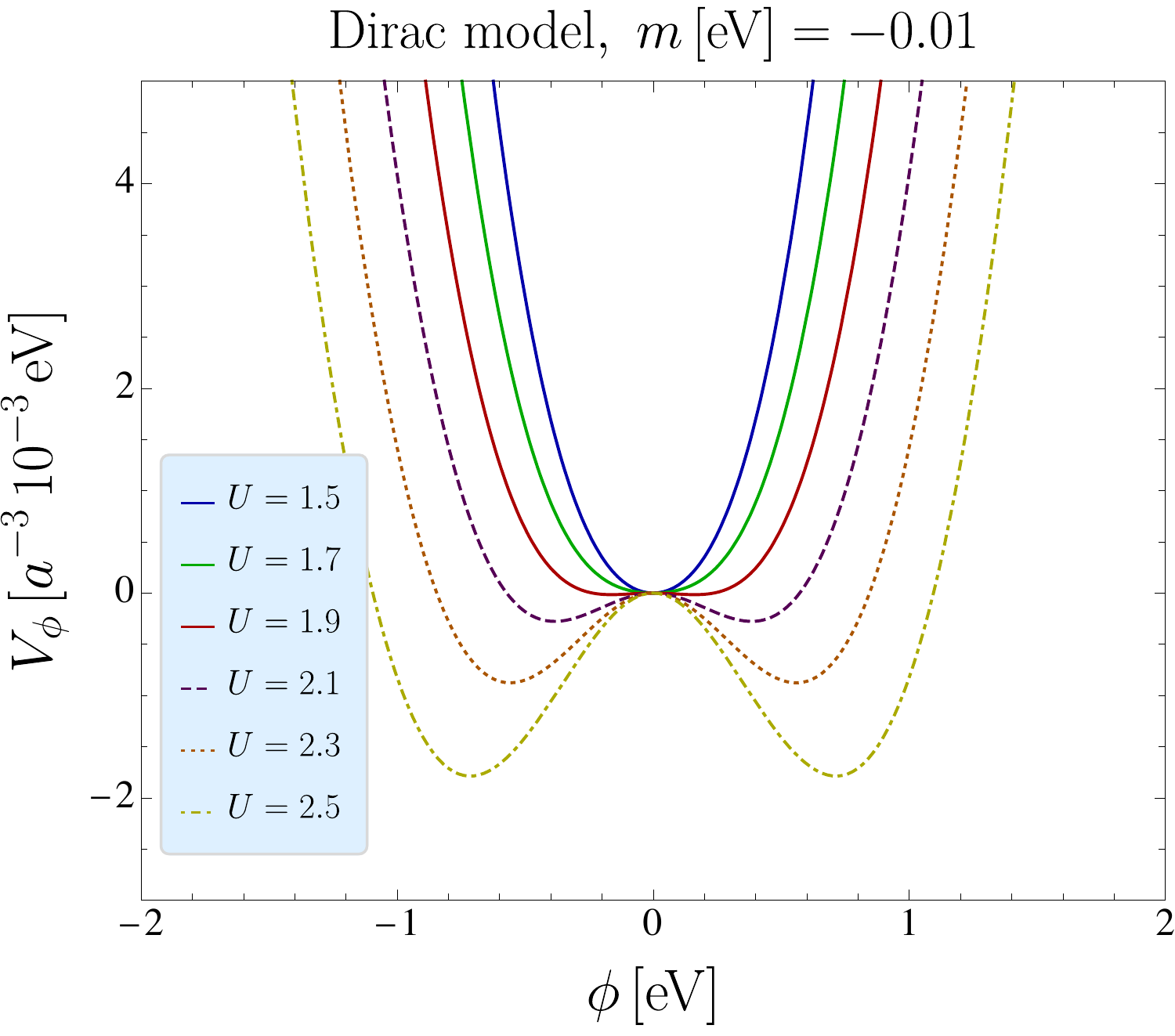}
    \includegraphics[scale=0.38]{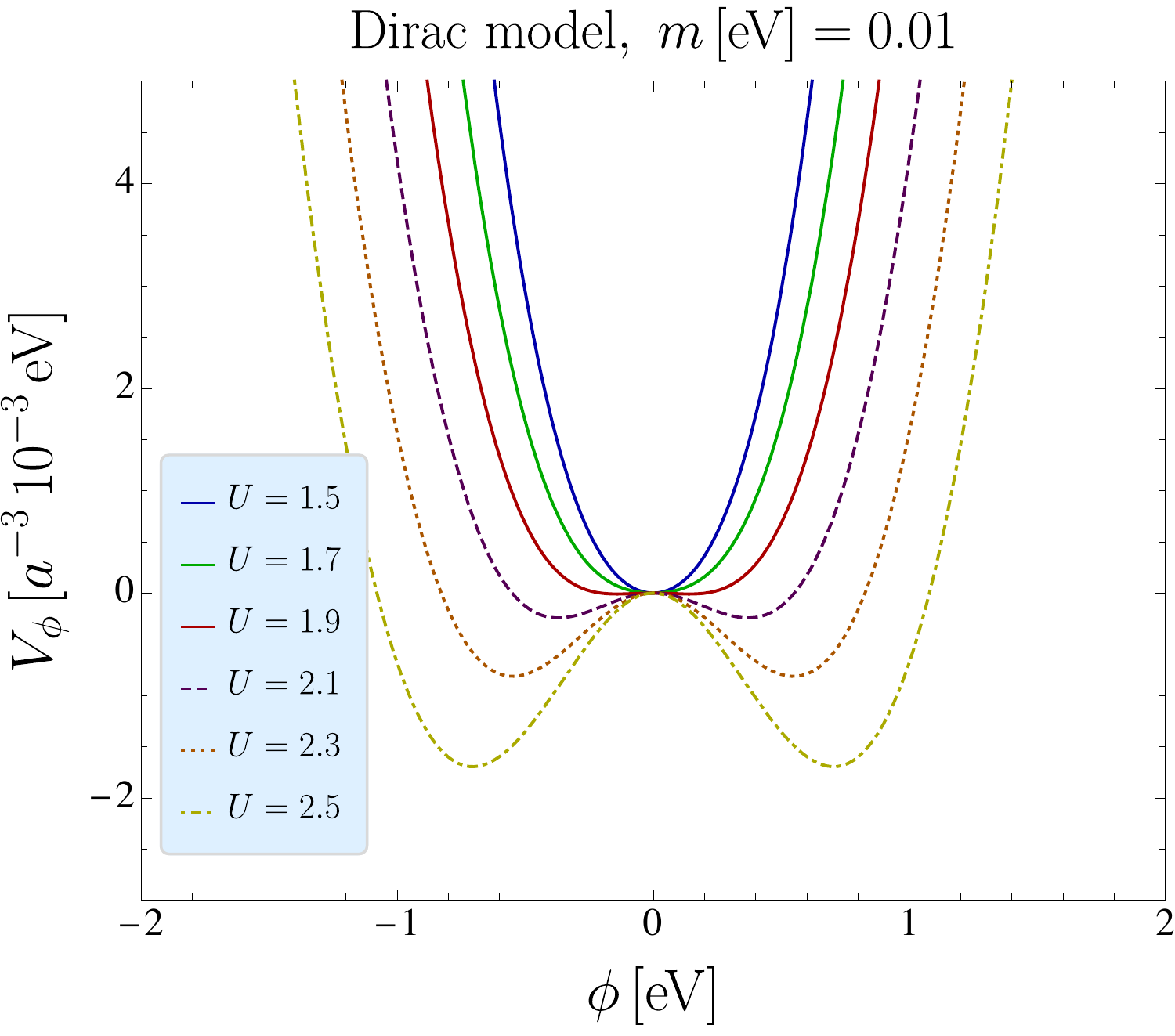}
    \includegraphics[scale=0.38]{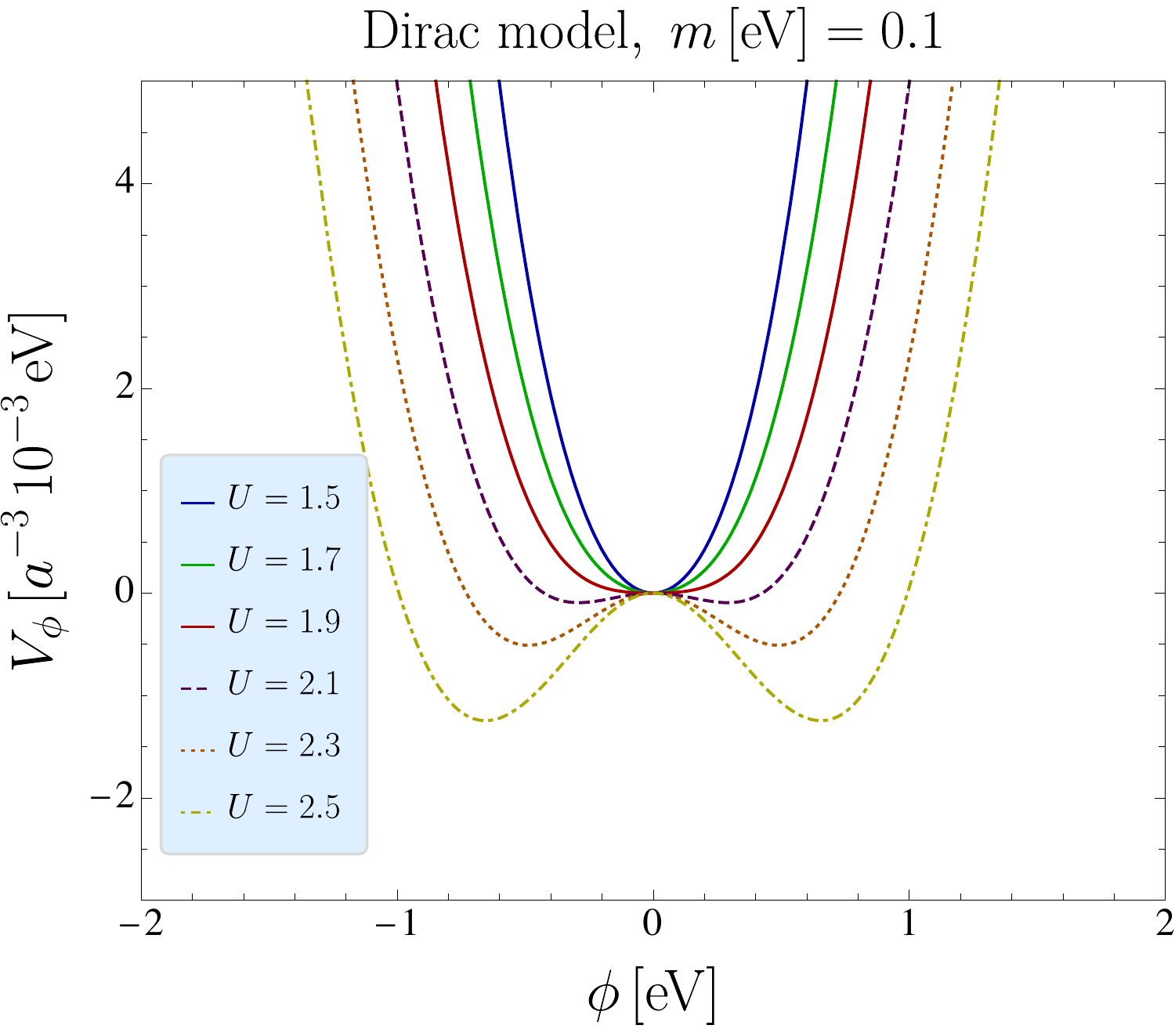}
    \includegraphics[scale=0.38]{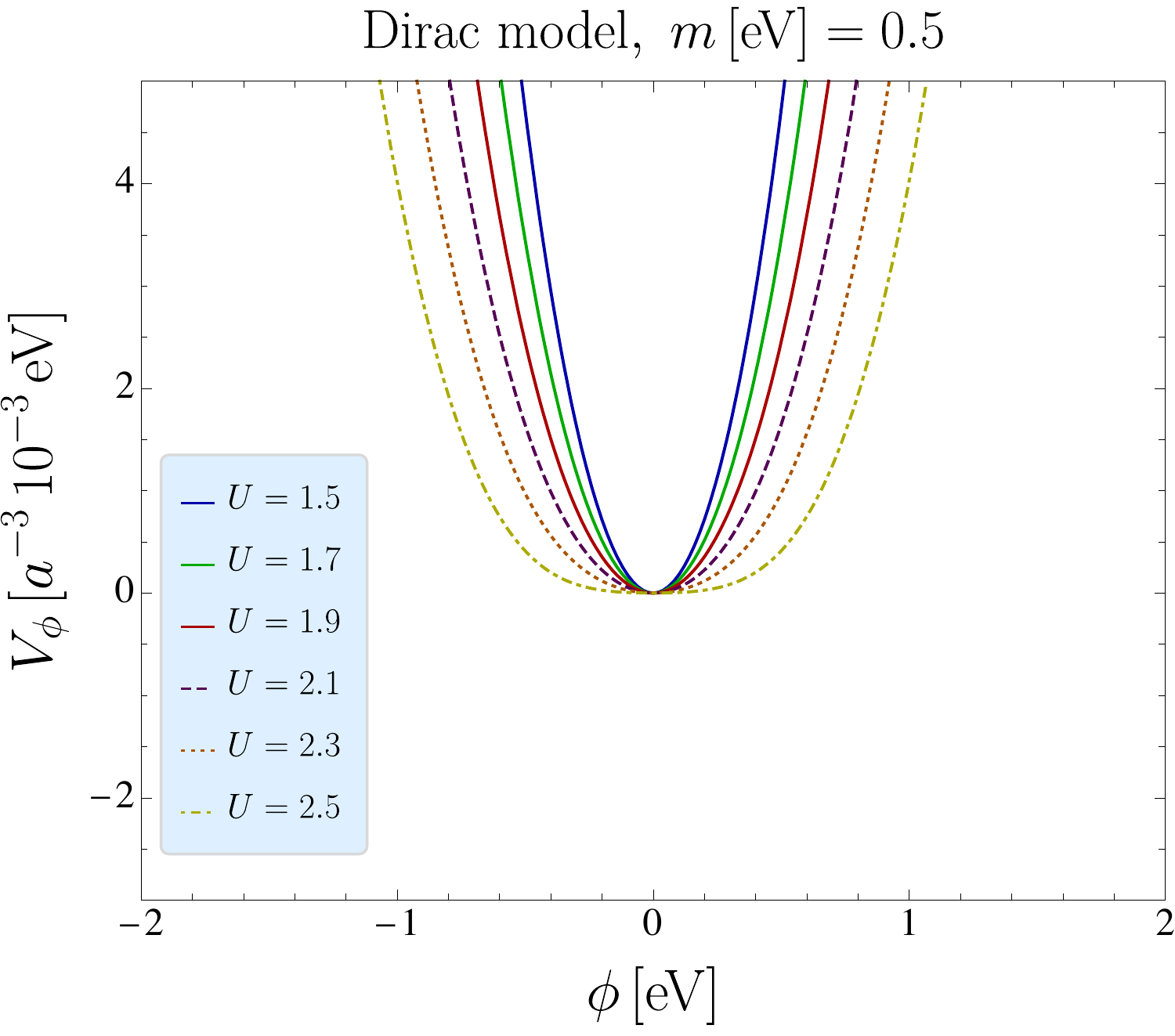}
  \end{center}
  \caption{$V_\phi$ as function of $\phi$ for various values of
    $m$~[eV] in Dirac model.  $A/a=1$~eV, $B/a^2=0.5$~eV is taken.  At
    each panel, $U$~[eV] is taken to 1.5 to 2.5 from top line to
    bottom line. We put a cutoff to the normalized momentum integral
    as $-1<\tilde{\vb*{k}}<1$.}
  \label{fig:JVphi_Dirac}
\end{figure*}

\begin{figure*}[t]
  \begin{center}

   \includegraphics[scale=0.35]{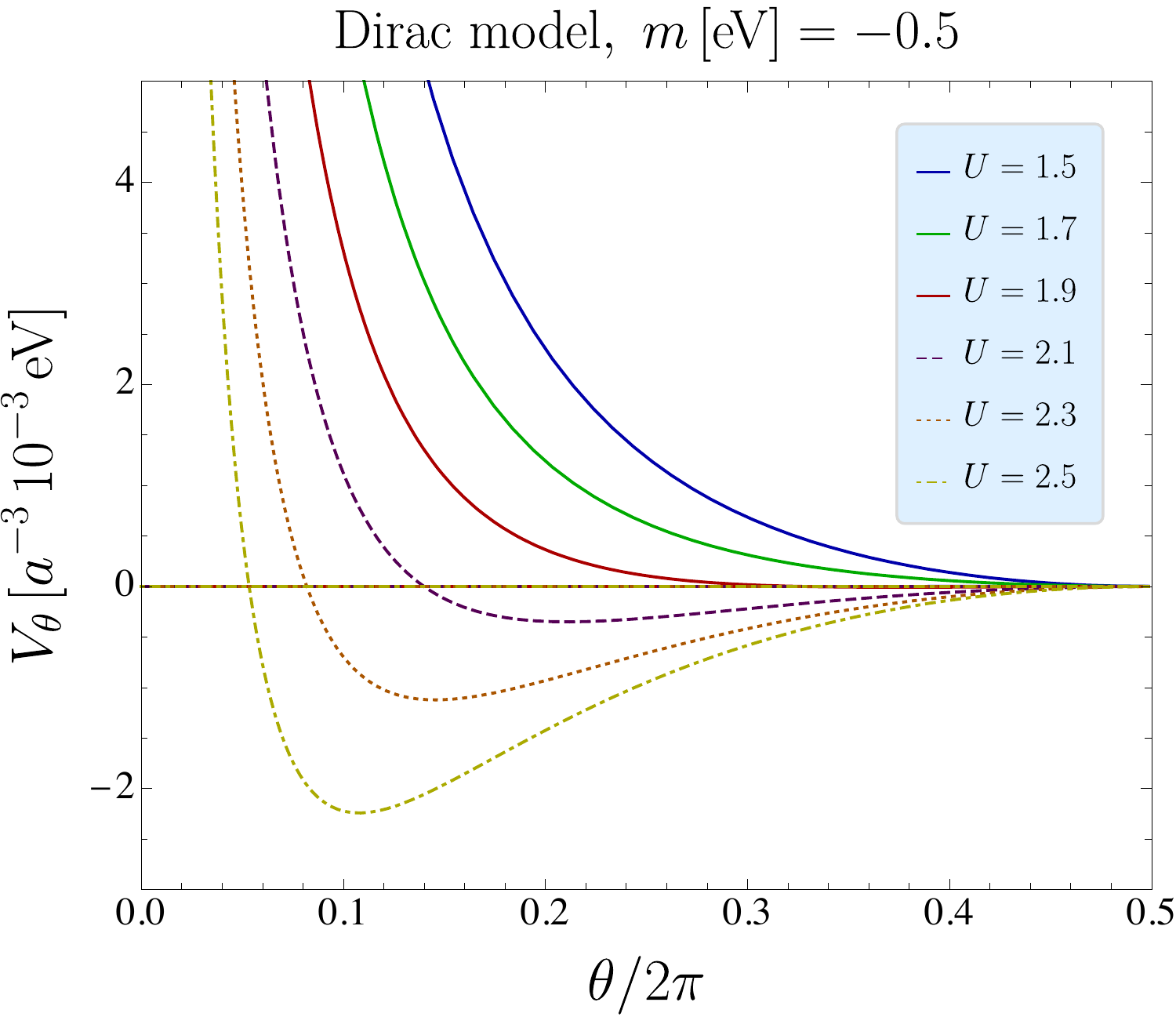}
    \includegraphics[scale=0.35]{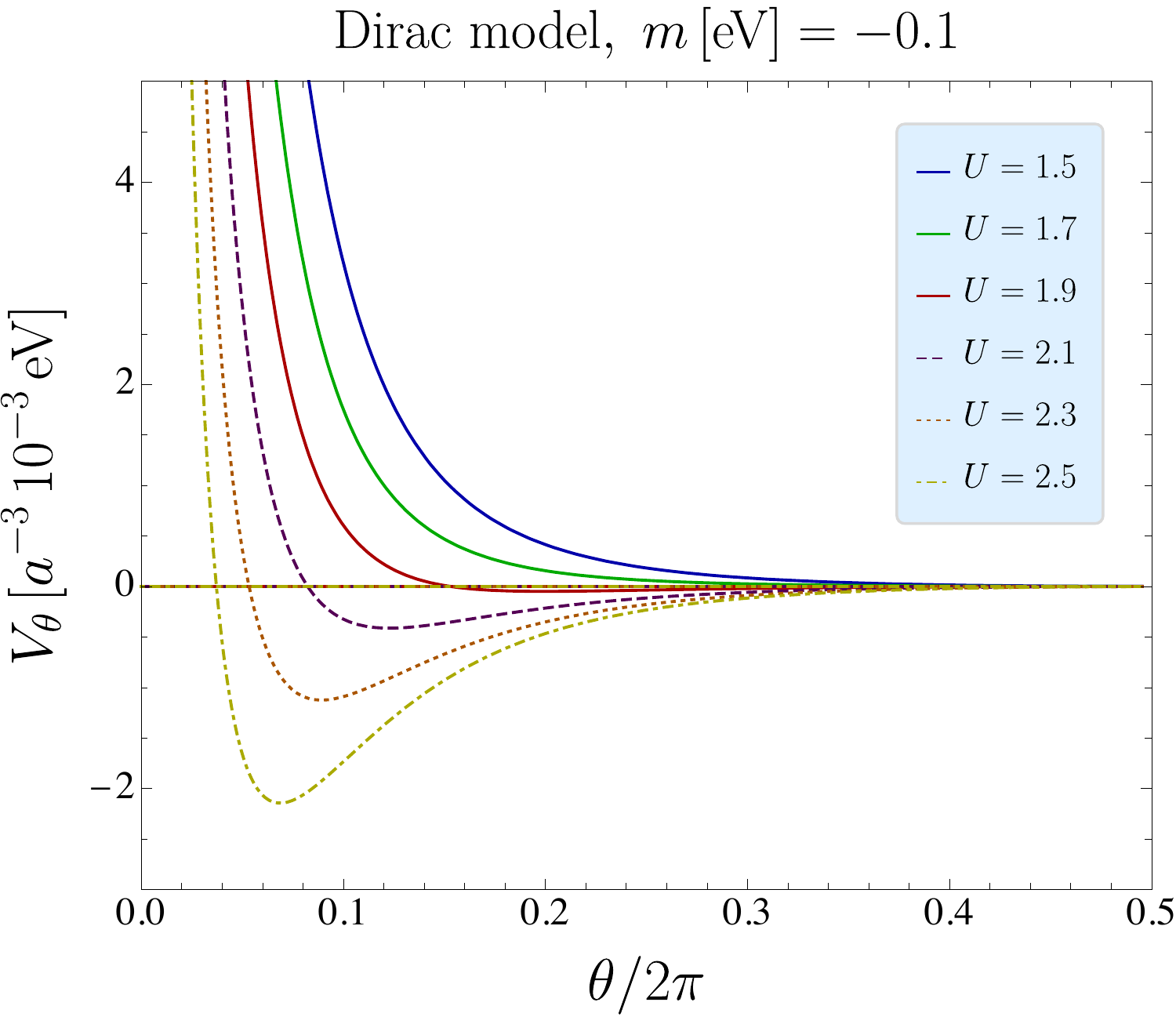}
    \includegraphics[scale=0.35]{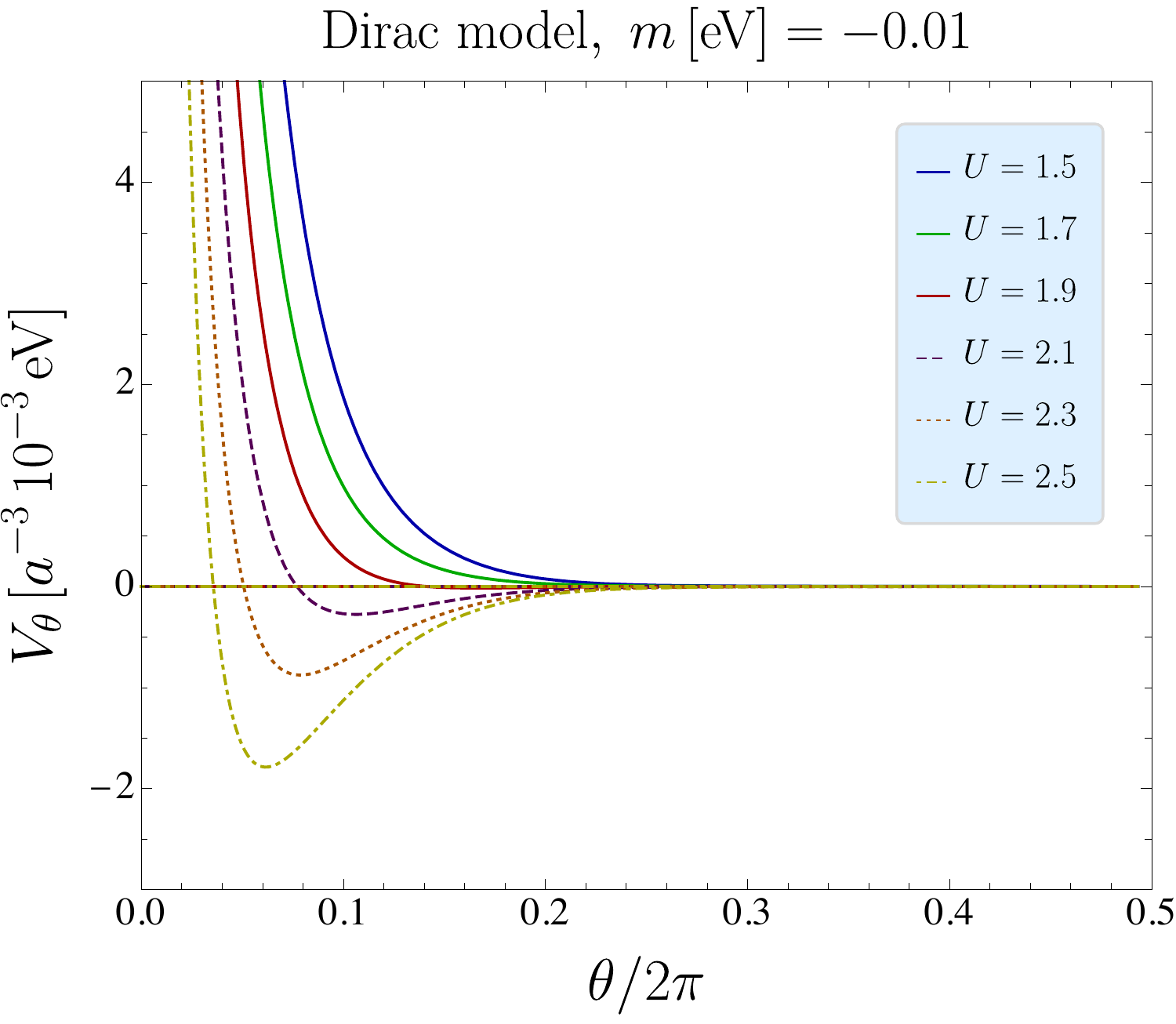}
    \includegraphics[scale=0.35]{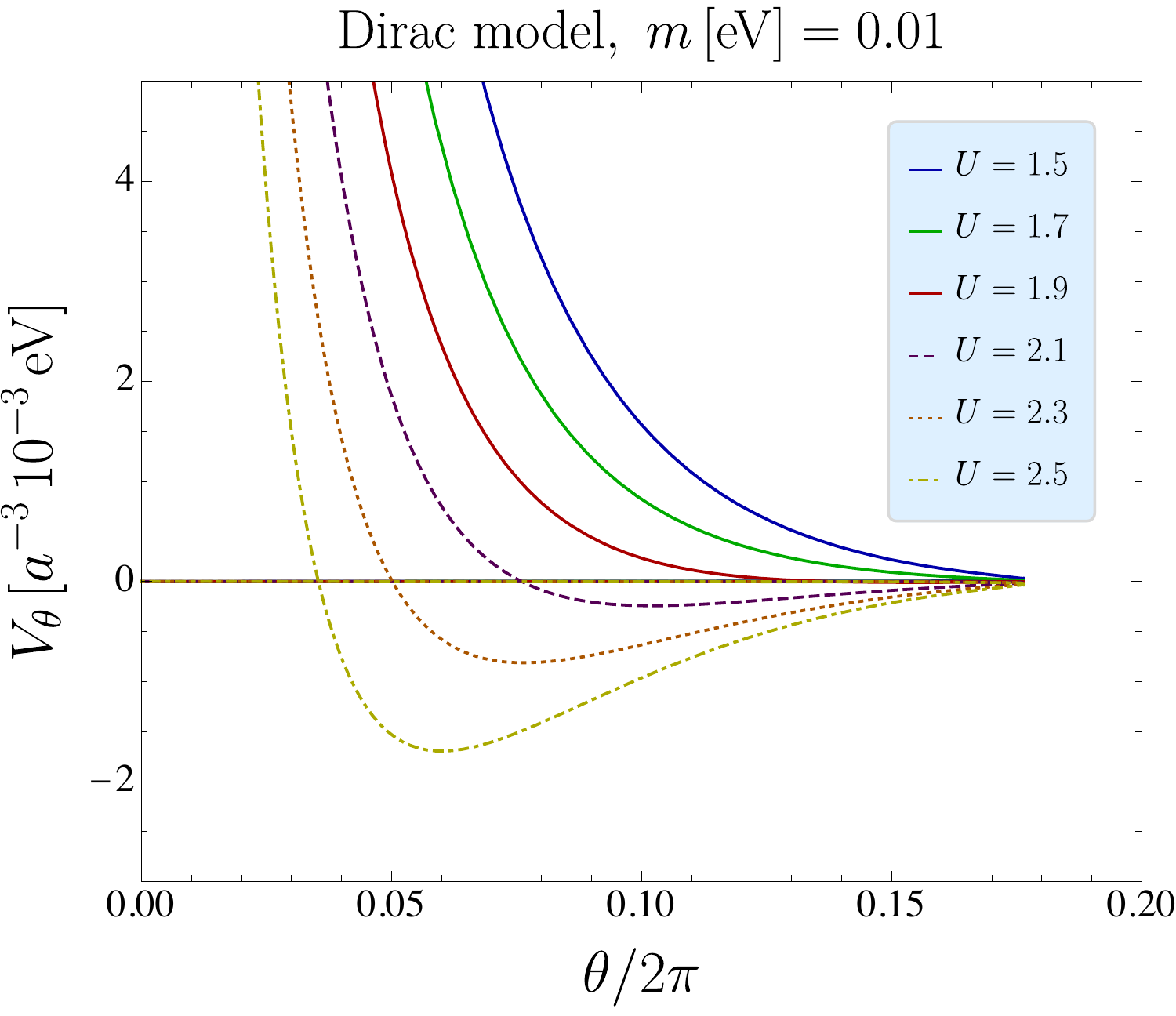}
    \includegraphics[scale=0.35]{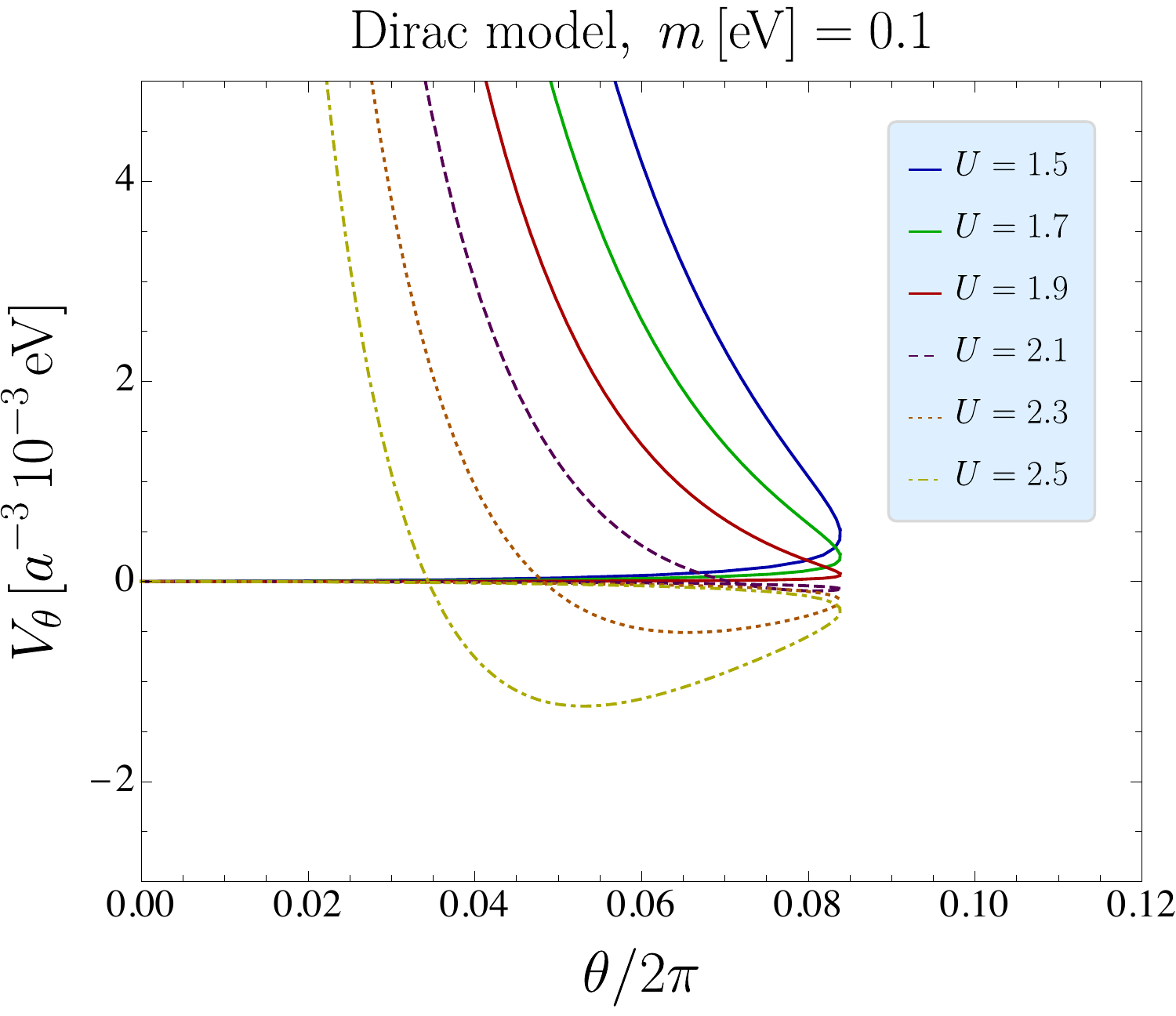}
    \includegraphics[scale=0.35]{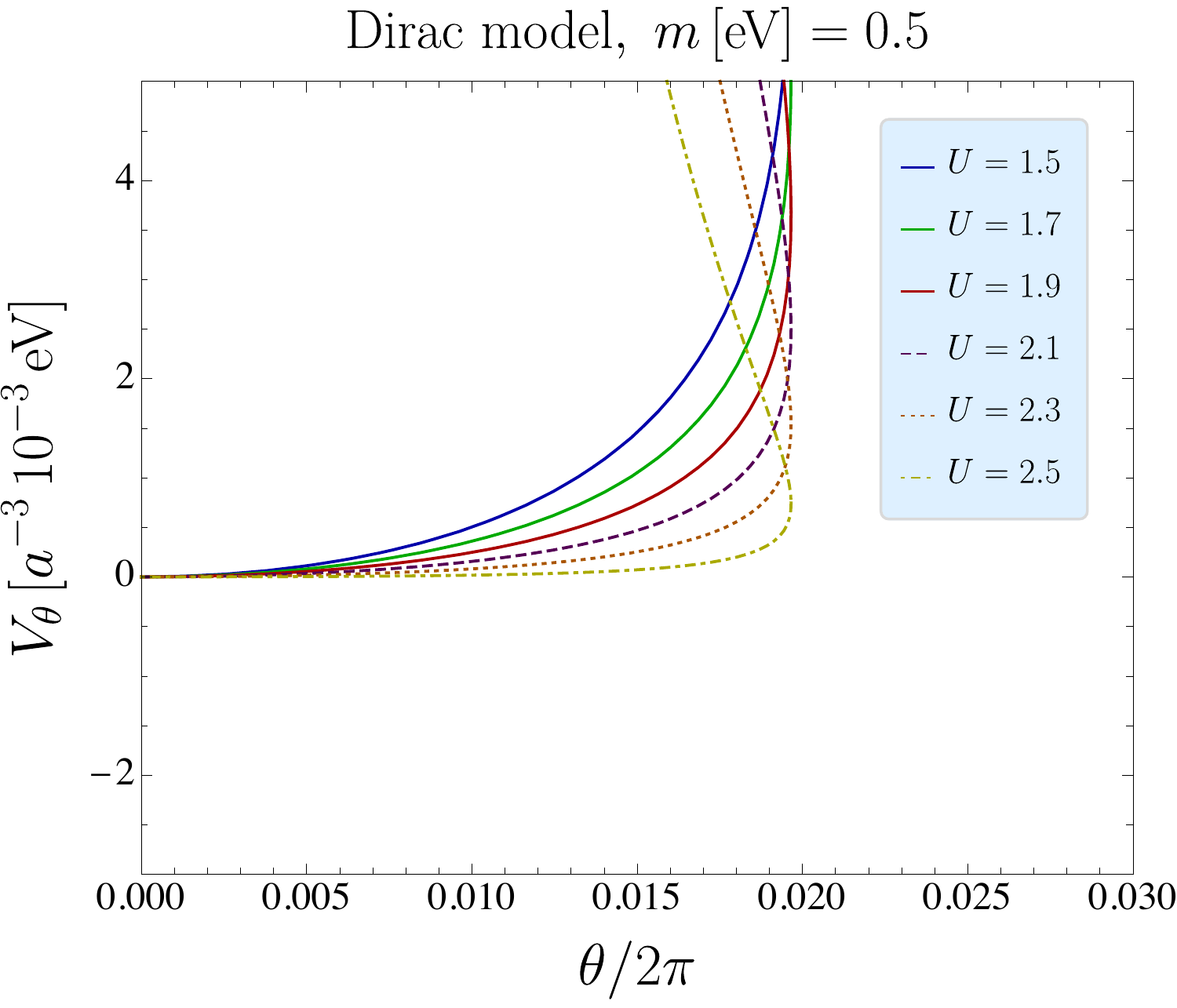}
  \end{center}
  \caption{$V_\theta$ as function of $\theta/(2\pi)$ for various
    values of $m$~[eV] in Dirac model. The parameters and line
    contents are the same as in Fig.\,\ref{fig:JVphi_Dirac}. }
  \label{fig:JVtheta_Dirac}
\end{figure*}


The effective potential for $\phi$ is derived as polynomial expansion from
Eq.\,\eqref{eq:Z_in_expand}. Using the Fourier transformation
of the propagator,
\begin{align}
  G_0(x)&=\int \frac{d^4k}{(2\pi)^4}\tilde{G}_0(k)e^{-ik\vdot x}\,,
  \\
  \tilde{G}_0(k&)=
  \frac{k^0-\epsilon_0(\vb*{k})+\sum_{a=1}^4d_0^a(\vb*{k})\Gamma^a}
       {(k^0-\epsilon_0(\vb*{k}))^2-|d_0(\vb*{k})|^2}\,,
      \label{eq:prop0}
\end{align}
where $k\vdot x=k^0x^0-\vb*{k}\vdot \vb*{x}$ and $|d_0|^2\equiv
\sum_{a=1}^4d^a_0d^a_0$, it is obtained as (see Appendix~\ref{sec:EL} for
detail)
\begin{align}
   V_\phi&=-2\int \frac{d^3k}{(2\pi)^3}
  (\sqrt{|d_0|^2+\phi^2}-|d_0|)
  +M^2\phi^2\,.
  \label{eq:Vphi}
\end{align}
A similar calculation is done in
Refs.\,\cite{Sekine:2020ixs,Schutte-Engel:2021bqm,Sekine:2015eaa}. However,
the expansion is truncated at $n=2$. On top of that, the mass term,
the second term in Eq.\,\eqref{eq:Vphi} is missing.\footnote{There are
other typos in Ref.\,\cite{Sekine:2015eaa}.  We thank A.~Sekine for
confirming this. } In fact, both the negative term derived from the
last term in Eq.\,\eqref{eq:Z_in_expand} and the mass term are crucial
to understand the AFM/FM phase and the topological/normal state, which will be
shown below.

Fig.\,\ref{fig:JVphi_Dirac} shows $V_\phi(\phi)$ as function of
$\phi$. In the calculation we introduce dimensionless wavenumber
$\tilde{k}\equiv ka$ where $a$ is the lattice size (typically \AA) and
take $A/a=1$~eV and $B/a^2=0.5$~eV\footnote{ A larger value of $B$ is
obtained in first-principles calculation~\cite{Zhang:2009zzf}. This
may indicate it is difficult to have the AFM order in the materials
such as Fe-doped Bi$_2$Te$_3$ (see also later discussion).  In our
analysis, we choose $\order{1}$~eV for $A$ and $B$. } and the integral
is executed in $[-1,\,1]$ region. This is because the Dirac model is
low energy effective model and the (normalized) momentum cannot take
as large as $\pi$.  (Instead if the effective model for 3D topological
insulators given in Ref.\,\cite{Li:2010} is used, then the momentum
integral is done in $[-\pi,\,\pi]$. See
Appendix~\ref{sec:results_in_Limodel} for the results.) It is seen
that the curvature at $\phi=0$ becomes negative when $U$ gets
larger. This situation can be understood by expanding $V_\phi$ around
$\phi=0$,
\begin{align}
  V_\phi=
 \int  \frac{d^3k}{(2\pi)^3}\Bigl[\frac{2}{U}-\frac{1}{|d_0|}\Bigr]\
  \phi^2 +{\cal O}(\phi^4)\,,
\end{align}
Here we have used Eq.\,\eqref{eq:M^2}.  It is seen that the
coefficient of $\phi^2$ term becomes negative when $U>U_{\rm crit}$
where $U_{\rm crit}$ satisfies
\begin{align}
  \int  \frac{d^3k}{(2\pi)^3}\left[\frac{2}{U_{\rm crit}}
    -\frac{1}{|d_0|}\right]
=0
  \,.
\end{align}
Namely, for $U<U_{\rm crit}$ the minimum of the potential located at
$\phi=0$, corresponding to the PM phase. When $U>U_{\rm crit}$, on the
other hand, the phase transition occurs to get nonzero $\phi_0$, which
satisfies
\begin{align}
  \int \frac{d^3k}{(2\pi)^3}
  \Bigl(\frac{2}{U}
    -\frac{1}{\sqrt{|d_0|^2+\phi_0^2}}
    \Bigr)=0\,,
\end{align}
and the AFM order appears. In the next section, we show how the
situation is described in terms of the axion field.

\section{Axion potential}
\label{sec:V_theta}

So far we have derived the effective potential for $\phi$. It can be
described by axion field because $\phi$ is related to the axion field
$\theta$. To be concrete, $\theta$ is given by $d^a$~\cite{Li:2010},
\begin{align}
  \theta = \frac{1}{4\pi} \int d^3k
  \frac{2|d|+d^4}{(|d|+d^4)^2|d|^3}
  \epsilon^{ijkl}d^i
  \partial_{k_x}d^j \partial_{k_y}d^k \partial_{k_z}d^l \,,
  \label{eq:theta_phi}
\end{align}
where $|d|^2=\sum_{a=1}^5d^ad^a$ and $\epsilon^{ijkl}$ is Levi-Civita
symbol with $i,j,k,l$ being 1, 2, 3, 5. It is checked that $\theta=0$
or $\pm \pi$ when $d^5=\phi=0$ and the value depends on $d^4$. (We
note that $\theta=+\pi$ and $-\pi$ are identical and hereafter we focus
on $\theta=\pi$ unless otherwise stated. See also
Appendix~\ref{sec:theta_chiral} for another derivation of $\theta$
using chiral anomaly.  However, it is not equivalent to the above
expression, except for a specific case.) For instance, in the Dirac
model given in Eq.\,\eqref{eq:DiracModel}, $m/B>0$ ($<0$) gives
$\theta=0$ ($\pi$) when $d^5=0$~\cite{Zhang:2019lkh}.  We have
confirmed that the results are consistent with
Ref.\,\cite{Zhang:2019lkh} (see Figs.\,\ref{fig:theta_phi_compare} and
\ref{fig:theta_m_compare} in Appendix~\ref{sec:results_in_Limodel}).
Using Eq.\,\eqref{eq:theta_phi}, we can derive $\phi$ as function of
$\theta$. Let us write
\begin{align}
  \phi = \Phi (\theta)\,.
\end{align}
Then, the potential as function of $\theta$ can be obtained by
\begin{align}
  V_\phi(\Phi(\theta)) \equiv V_\theta(\theta)\,.
\end{align}

\begin{figure}[t]
  \begin{center}
    \includegraphics[scale=0.5]{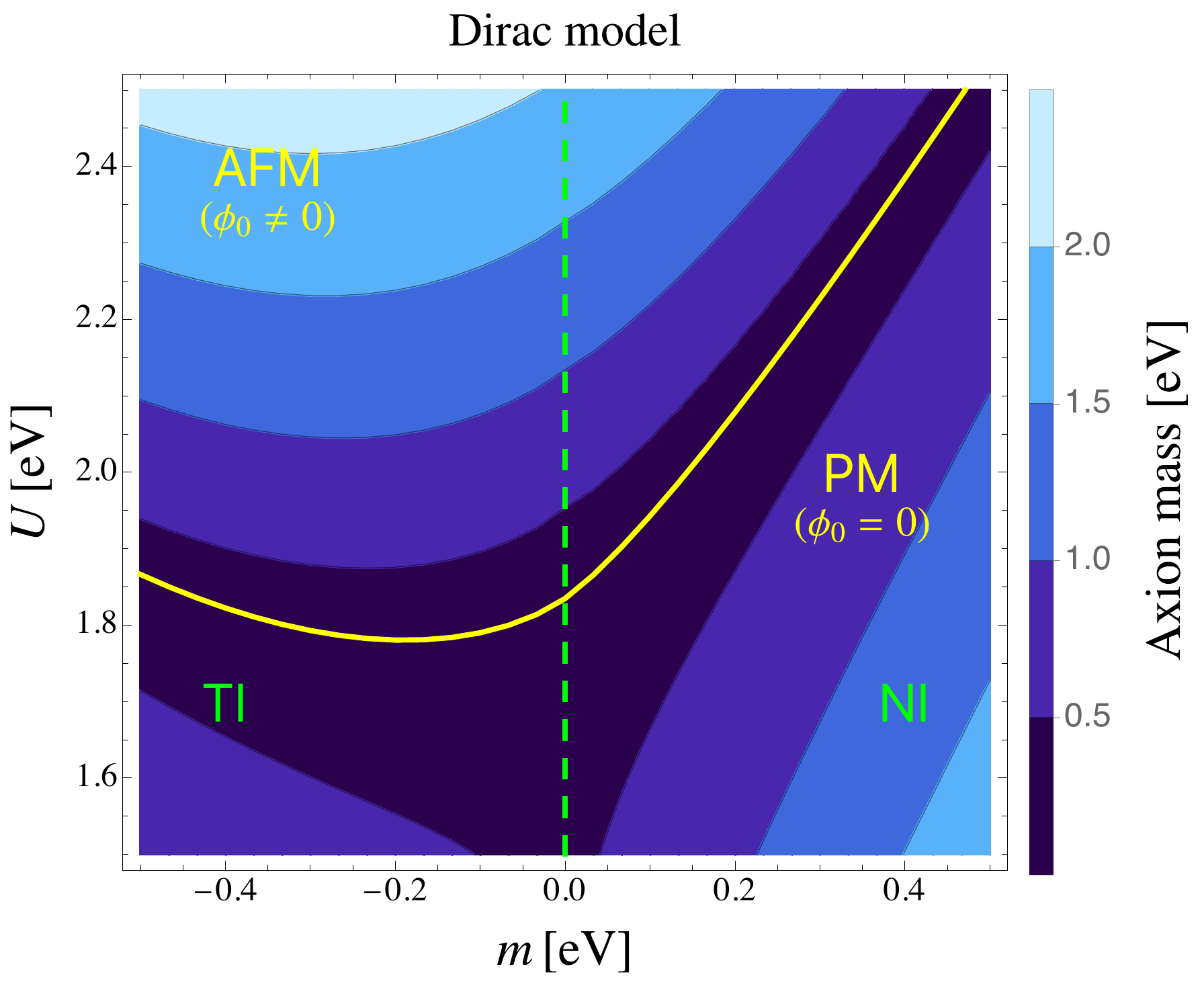}
      \end{center}
  \caption{Axion mass on ($U$,\,$m$) plane in the Dirac model.
    $A/a=1$~eV, $B/a^2=0.5$~eV is taken.  Solid (yellow) line shows
    the critical points. Upper and lower regions separated by the
    solid line correspond to antiferromagnetic order (``AFM'') and
    paramagnetic order (``PM''), respectively. Dotted (green) line is
    boundary of topological insulator phase (``TI'') and normal
    insulator phase (``NI'').  }
  \label{fig:ma}
\end{figure}

Fig.\,\ref{fig:JVtheta_Dirac} shows $V_\theta$ as function of
$\theta/(2\pi)$ in the Dirac model. It is seen the axion potential
nontrivially depends on the sign of $m$. When $U$ is smaller than
$U_{\rm crit}$, the potential minimum is located at $\theta=0$ $(\pi)$
for $m>0$ $(m<0)$. $V_\theta$ is multivalued function of $\theta$
since $\phi$ is multivalued function of $\theta$ for $m>0$. For $m<0$,
on the other hand, $\theta$ is monovalent function, {\it i.e.},
$\theta\to \pm \pi$ for $\phi \to 0\pm$.  (See
Fig.\,\ref{fig:theta_phi_compare} in Appendix
\ref{sec:results_in_Limodel}.)  In terms of potential minimum of
$V_\phi$, both $\theta=0$ and $\pi$ correspond to $\phi$ (or $m_5$)
$=0$, {\it i.e.}, the PM phase. In terms of $\theta$, on the other
hand, they are different topologically. Namely, $\theta=0$ ($\pi$) is
classified as the normal insulator (topological insulator). When $U$
gets larger than $U_{\rm crit}$, nonzero $\phi_0$ becomes a minimum of
the potential and insulator is in the AFM order. In terms of axion,
both $\theta=0$ and $\pi$ becomes unstable and a new stable point
appears. It should be noticed that the effective potential given in
Eq.\,\eqref{eq:Vphi} is crucial to derive this result, which consists
of the negative potential term derived from integrating out the
electron field and the positive mass term that comes consequently from
the Hubbard-Stratonovich transformation of the Hubbard interaction
term \eqref{eq:H_int}.

Now the definition of {\it static} and {\it dynamical} axions are
clear. The static axion $\theta_0$ is defined as a stationary point of
$V_\theta(\theta)$, and the dynamical axion $a$ as a fluctuation
around $\theta_0$. Denoting $\phi_0$ as corresponding value of
$\theta_0$, we expand $\phi$ and the potential $V_\phi$,
\begin{align}
  \phi &=  \phi_0+\dv{\Phi(\theta)}{\theta}\eval_{\theta=\theta_0}a+\cdots\,,
  \label{eq:phi_expand}
  \nonumber \\
  &\equiv \phi_0+g a+\cdots\,,
  \\
  V_\phi &= V_\phi(\phi_0)+
  \frac{1}{2}g^2\left[\dv[2]{V_\phi}{\phi}\right]_{\phi=\phi_0}a^2+
  \cdots
  \nonumber \\
  &\equiv
  V_\phi(\phi_0)+
  Jg^2m_a^2a^2+\cdots\,,
  \label{eq:V_phi_expand}
\end{align}
where $J$ is called stiffness and obtained as (see
Appendix~\ref{sec:EL} for derivation)
\begin{align}
  J=
  \int  \frac{d^3k}{(2\pi)^3}\frac{|d_0|^2}{4(|d_0|^2+\phi_0^2)^{5/2}}\,.
  \label{eq:J}
\end{align}
Using Eqs.\,\eqref{eq:phi_expand} and \eqref{eq:V_phi_expand}, the
action for the dynamical axion $a$ is obtained,
\begin{align}
  S_a=Jg^2\int \dd[4]{x}
  a
  \left[-\partial_t^2+(\vb*{v}\cdot \nabla)^2-m_a^2\right]a+{\cal O}(a^4)
  \,.
\end{align}
Therefore, the axion mass is obtained by
\begin{align}
  Jm_a^2=
  \int  \frac{d^3k}{(2\pi)^3}\left[\frac{2}{U}-\frac{1}{|d_0|}\right]\,,
  \label{eq:Jma2_1}
\end{align}
for $U<U_{\rm crit}$ and 
\begin{align}
  Jm_a^2=
  \int\frac{d^3k}{(2\pi)^3}
  \frac{\phi_0^2}{(|d_0|^2+\phi_0^2)^{3/2}}\,,
  \label{eq:Jma2_2}
\end{align}
for $U>U_{\rm crit}$. Eqs.\,\eqref{eq:J} and \eqref{eq:Jma2_2} are
consistent with the results in Ref.\,\cite{Li:2010}, except for a
factor of 4 larger compared to ones in the literature. In the
literature, however, how the value of $\phi_0$ is determined is not
clearly stated. Rather it is treated as another parameter that is
independent of the model parameter in $d^a$.  By using the effective
potential, on the contrary, $\phi_0$ is determined uniquely as a
stationary point of the potential when the model parameters are fixed.

To demonstrate the result, we plot the axion mass on phase plane in
Fig.\,\ref{fig:ma}.  There are phases classified by the AFM/PM and the
topological/normal orders, and they are determined by $U$ and $m$ for
the given parameters $A$ and $B$ of the Dirac model. The AFM (PM) phase
corresponds to $\phi_0=$nonzero ($0$), and the topological (normal)
phase corresponds to $m<0$ ($m>0$). In the PM phase, $\theta_0 =0$
($\pi$) in the normal (topological) phase meanwhile $\theta_0$ takes a
value of $0<\theta_0<\pi$ in the AFM phase depending on $\phi_0$. It
is found that the axion mass is typically $\order{1}$ eV in the four
phases. (The same result is obtained in the effective model for 3D
topological insulators. See Fig.\,\ref{fig:ma_Li} in Appendix
\ref{sec:results_in_Limodel}.) This result seems to be different from
the value argued in the literature~\cite{Li:2010} where the $\order{\rm
  meV}$ axion in the topological AFM insulators is discussed. Let us
elaborate on this issue closely.

First of all, it is clear that the existence of dynamical axion does
not depend on whether insulator is topological or not. This point is
mentioned in Ref.\,\cite{Ooguri:2011aa}. Fig.\,\ref{fig:ma} shows that
the axion mass can be much suppressed near the critical point of the
AFM phase in both the topological and normal insulators. In
Ref.\,\cite{Li:2010}, on the other hand, $m_5\,(=\phi_0)\,=1$ meV is
taken, which leads to meV scale axion mass.  At a glance this looks
possible; a small value of $U$ can give a suppressed $m_5$ since
$m_5\propto U M_z^-$ in mean field approximation ($M_z^-$ in
Ref.\,\cite{Li:2010} corresponds to $m^z$ in
Appendix~\ref{sec:MFA}). However, it is no longer the AFM order if
$U<U_{\rm crit}$. Therefore, suppressed $U$ cannot be naively
considered in this simple setting. This relates to an intrinsic issue
that ${\rm Bi}_2{\rm Se}_3$ does not have an AFM phase. A possible way
out suggested in Ref.\,\cite{Li:2010} is doping ${\rm Fe}$ to create
the AFM order. In this way, the AFM order could be realized. However,
doping Fe may not be successful due to the first-principles
calculation~\cite{Zhang:2013jz}. A current candidate is Mn-doped
topological insulator, Mn$_2$Bi$_2$Te$_5$ suggested in
Ref.\,\cite{Zhang:2019lkh}. The first-principles calculation done by
Ref.\,\cite{Li:2020fvr} shows that Mn$_2$Bi$_2$Te$_5$ has rich
magnetic topological quantum states, including the AFM, FM and other
magnetic states. The energy gap of the AFM phase indicates $m_5\sim
{\cal O}(10\,\mathchar`-\,100~{\rm meV})$. The total energy of each
magnetic states are also found to be the same order, which might
indicate that controlling the realization of the magnetic states would
be challenging. Therefore, the next step is to confirm these magnetic
states experimentally.\footnote{Another interesting candidate is
MnBi$_2$Te$_4$-family materials. Although they are not supposed to
have dynamical axion, they are found to have controllable AFM and
ferromagnetic orders. Therefore it is worth studying their properties
for the future application. See, for example,
Refs.\,\cite{Li:2019j,Li:2019h,Li:2019jia,Zhang:2019don,Yue:2019,Hao:2019}.}

In a nutshell, the axion mass is roughly $\order{\rm eV}$ but it can
be suppressed near the critical boundary between the AFM and PM orders.
Such dynamical axion with various mass range might be utilized in the
future {\it particle} axion or ALPs search.  On the other hand, in
order to precisely predict the axion mass in such specific
circumstances, the first-principles calculation to determine the
parameters is required.  Besides, we need experimental evidence of
having the AFM order in Fe- or Mn-doped Bismuth Selenide or Bismuth Telluride.

It is worth noticing that the dynamical axion should exist in the PM
phase.\footnote{Ref.\,\cite{Wang:2015hhf} pointed out the existence of
dynamical axion in topological PM insulators.}  On the other hand,
there would be other degree of freedom in the PM phase. For example,
$\phi$ is introduced in the circumstance where the AFM order is assumed
(see, {\it e.g.}, Eq.\,\eqref{eq:AssumingAFM}). If this assumption is
relaxed, another scalar field should appear. Therefore, it would be
nontrivial to find out dynamical axion mode under such contaminated
circumstances. If the possible obstacles are overcome, however, the
dynamical axion in the PM phase might be another tool for the future {\it
  particle} axion search.

\section{conclusions and discussion}
\label{sec:conclusions}

We have investigated axion in AFM insulators.  In the analysis, the
effective potential for the order parameter of the AFM phase is
derived from the path integral. From the effective potential, the
order parameter of the AFM is determined as the stationary point of
the potential, which distinguishes the AFM or PM order.  Axion
potential is obtained from the effective potential. It is shown that
the axion field value $\theta=0$ or $\pi$ is the minimum of the axion
potential in the PM order, meanwhile in the AFM order different values
of $\theta$ are obtained as the minimum of the potential. The field
value of the stationary point of axion is the {\it static} axion, and
the {\it dynamical} axion is defined as a fluctuation around
stationary value. The dynamical axion is predicted in four types of
phases distinguished by the AFM/PM and the topological/normal
states. The mass of axion turns out to be typically $\order{\rm
  eV}$. On the other hand, it can be much suppressed near the phase
boundary between the AFM and PM orders.  In either case, it is crucial
to determine the value of the order parameter in a possible candidate,
{\it e.g.}, Mn$_2$Bi$_2$Te$_5$, to predict the axion mass.

Another possible direction to investigate the static/dynamical axion
is to utilize the magnetoelectric effect. The magnetoelectric
couplings in generic insulators are obtained by the first-principles
calculation by Ref.\,\cite{Coh:2011}. For example, $\theta=1.3\times
10^{-3}$ and $1.07\pi$ are obtained for Cr$_2$O$_3$ and
Bi$_2$Se$_3$. In addition, if Bi is magnetized, the value of $\theta$
is shifted to $\theta=\pi \pm 0.55$. On the experimental side,
quantized Faraday and Kerr rotation due to Bi$_2$Se$_3$ film is
observed precisely~\cite{Wu:2016oxw}. Recently, anisotropic
topological magnetoelectric effect is observed in axion
insulators~\cite{Liu:2020wts}. Those are examples of ``axion
electrodynamics'' as discussed in Refs.\,\cite{Li:2010,Ooguri:2011aa}.
Such methods using optical responses would make it possible to
directly probe the order parameter of the magnetic states and the
properties of the static/dynamical axion.  The formalism given here
would be a help for finding and modeling realistic materials for the
future {\it particle} axion or ALPs search. Searching a wide range of
materials including ones that have no topological states or have
ferromagnetic/paramagnetic phases might open a new possibility for
{\it particle} axion search.

\begin{acknowledgments}
  We are grateful to Makoto Naka for valuable discussions in the early
  stage of this project and careful reading of the manuscript. We also
  thank Akihiko Sekine for useful discussions. This work was supported
  by JSPS KAKENHI Grant Numbers JP17K14278, JP17H02875, JP18H05542, and
  JP20H01894.
\end{acknowledgments}




\begin{widetext}

\appendix

\section{Effective model for 3D topological insulators}
\label{sec:results_in_Limodel}

We give numerical results in the effective model for 3D topological
insulators studied in Ref.\,\cite{Li:2010}.  The model is parametrized
as,
\begin{align}
   (d^1,\,d^2,\,d^3,\,d^4,\,d^5)
  &=(A_2\sin k_x,\,A_2\sin k_y,\,A_1 \sin k_z,\,{\cal M},m_5)\,,
  \nonumber \\
  {\cal M}
  &=M-2B_1-4B_2+2B_1 \cos k_z +2B_2 (\cos k_x + \cos k_y)\,,
  \label{eq:LiModel}
\end{align}
where the gamma matrices are the same as Eq.\,\eqref{eq:Gammas}.  If
we take $A_1=A_2=A$, $B_1=B_2=-B$, and $M=m$, and expand around
$\vb*{k}=0$, this model reduces to the Dirac model given in
Eq.\,\eqref{eq:DiracModel}. Fig.\,\ref{fig:theta_phi_compare}
(\ref{fig:theta_m_compare}) shows $\theta$ as function of $\phi$ ($M$
or $m$). As comparison, the results in the Dirac model are also
plotted. It is seen that the results are qualitatively the same as the
Dirac model.

Similar to Figs.\,\ref{fig:JVphi_Dirac}, \ref{fig:JVtheta_Dirac}, and
\ref{fig:ma}, $V_\phi$, $V_\theta$, and the axion mass computed in the
effective model for 3D topological insulators are plotted in
Figs.\,\ref{fig:JVphi_Li}, \ref{fig:JVtheta_Li}, and \ref{fig:ma_Li},
respectively.

\begin{figure}[t]
  \begin{center}
    \includegraphics[scale=0.45]{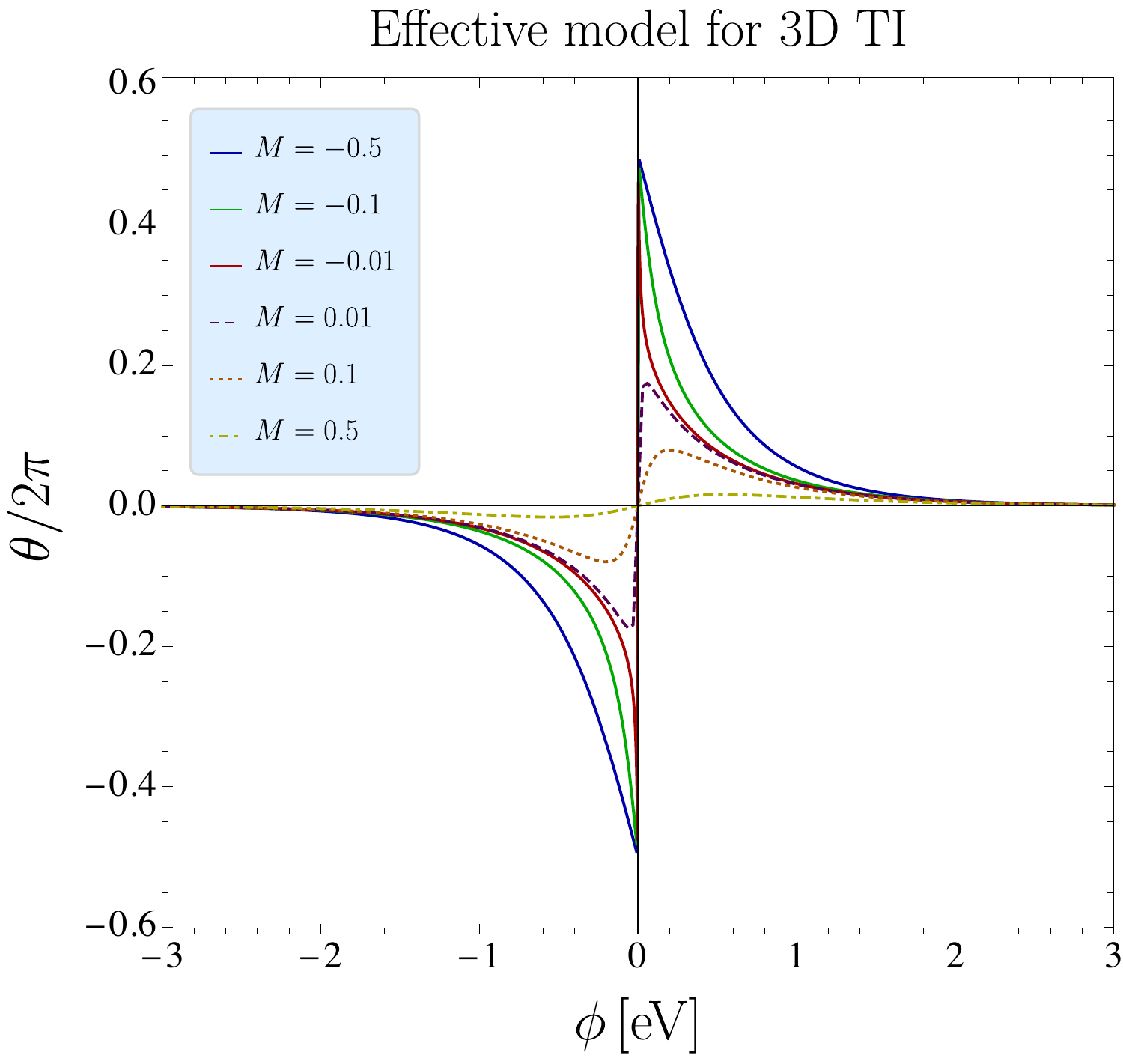}
    \includegraphics[scale=0.45]{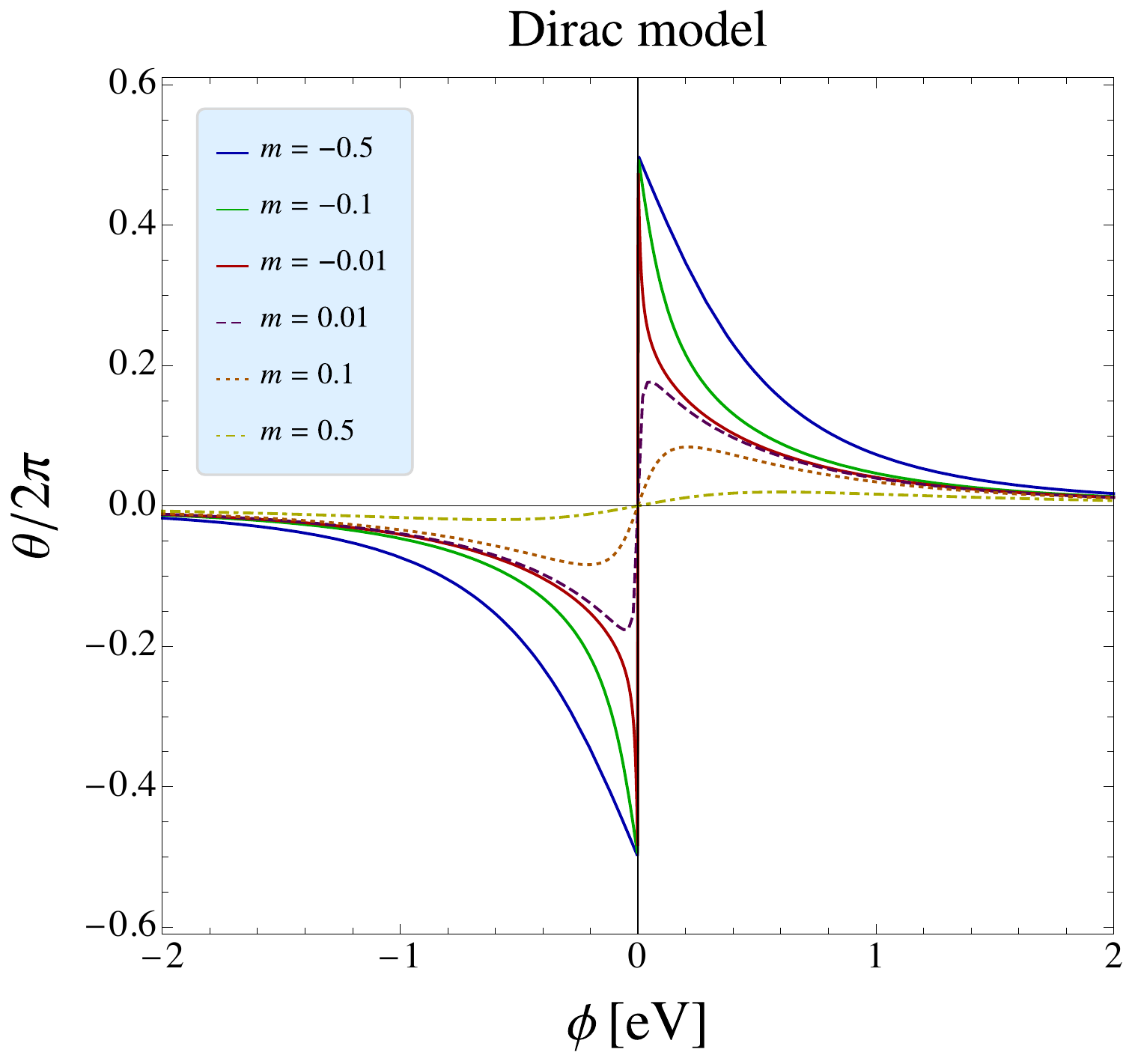}    
  \end{center}
  \caption{$\theta$ as function of $\phi$ for $M\,{[\rm eV]}=\pm
    0.01,$ $\pm 0.1,$ and $\pm 0.5$ in the effective model for 3D
    topological insulators with $A_1/a=A_2/a=1$~eV and
    $B_1/a^2=B_2/a^2=-0.5$~eV (left) and in the Dirac model given with
    $m=M$, $A=1$~eV, and $B=0.5$~eV (right).}
  \label{fig:theta_phi_compare}
\end{figure}

\begin{figure}[t]
  \begin{center}
    \includegraphics[scale=0.45]{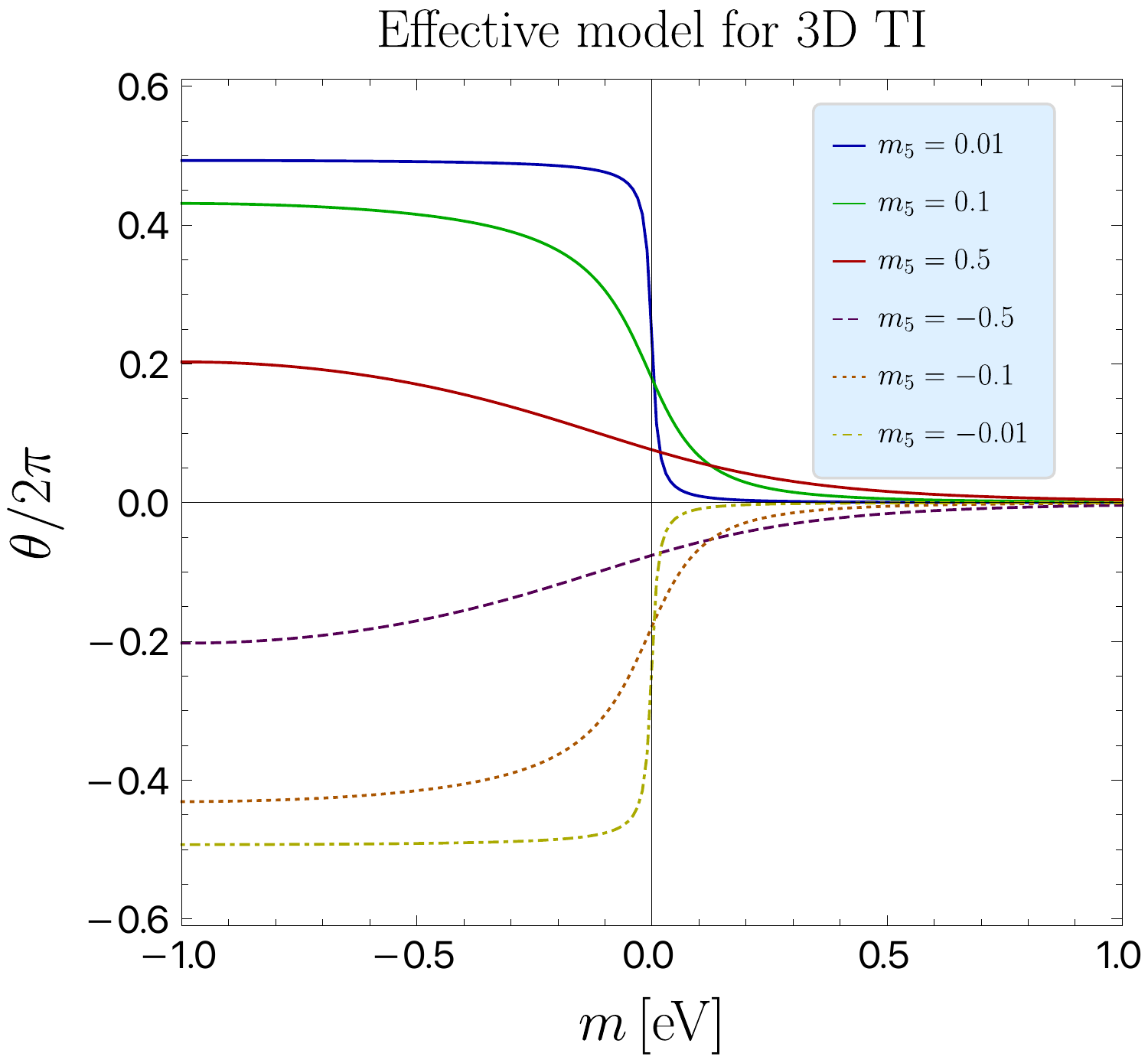}
    \includegraphics[scale=0.45]{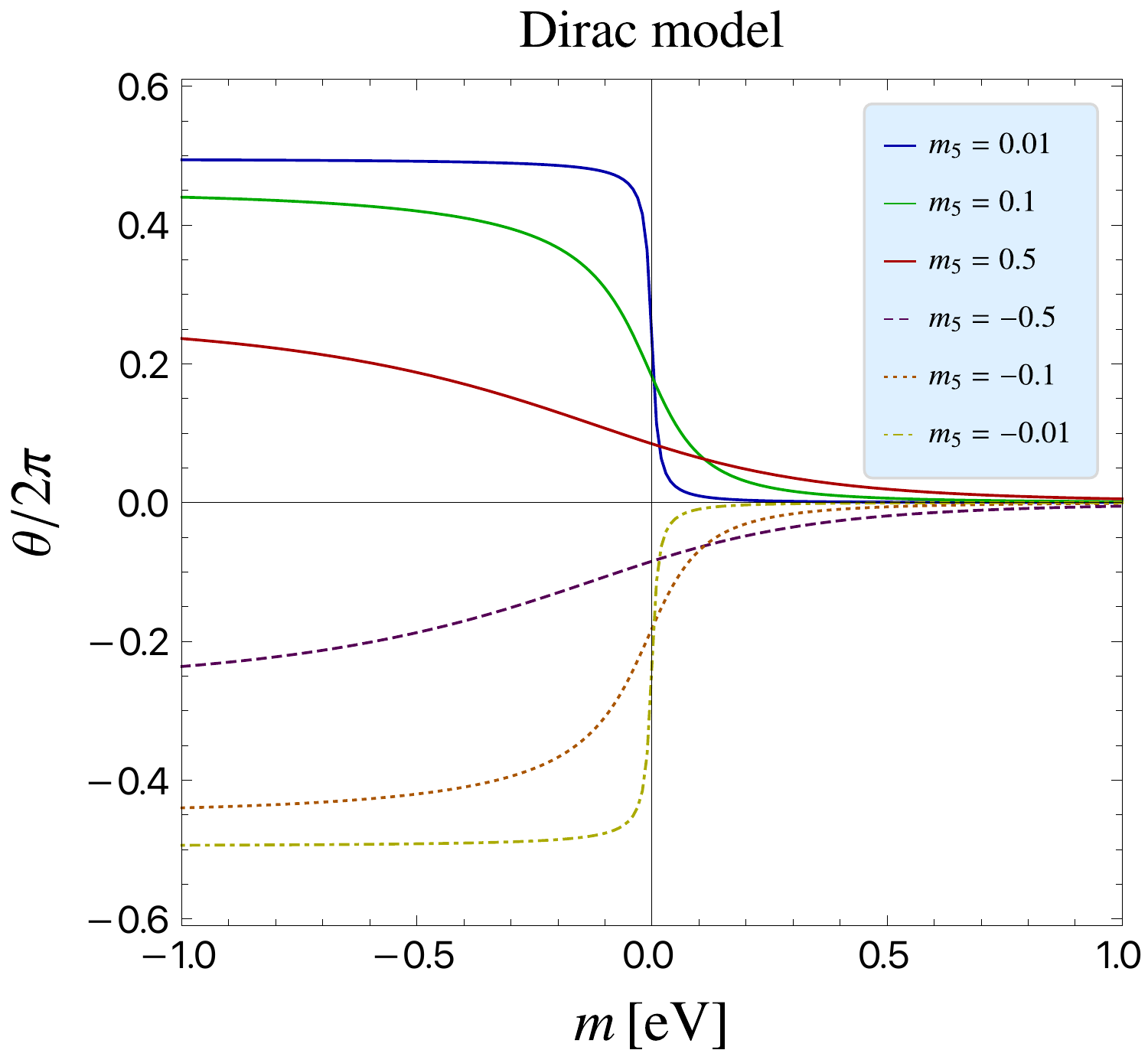}
  \end{center}
  \caption{$\theta$ as function of $M$ (or $m$) for $m_5\,[{\rm eV}]=\pm
    0.01,$ $\pm 0.1,$ and $\pm 0.5$ in the effective model for 3D
    topological insulators (left) and in the Dirac model (right). The
    other parameters are the same as
    Fig.\,\ref{fig:theta_phi_compare}.}
  
  \label{fig:theta_m_compare}
\end{figure}

\begin{figure}[t]
  \begin{center}
    \includegraphics[scale=0.38]{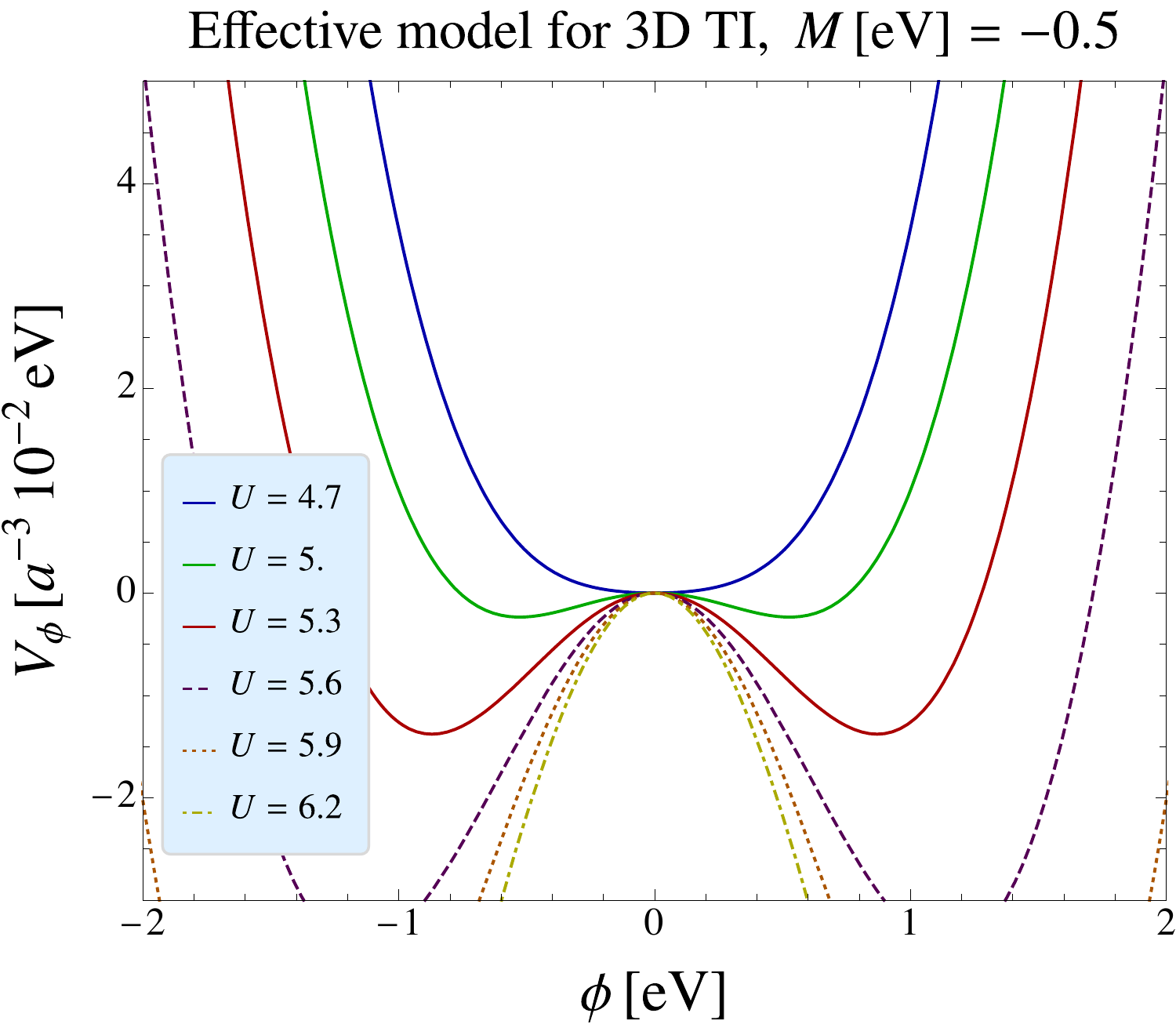}
    \includegraphics[scale=0.38]{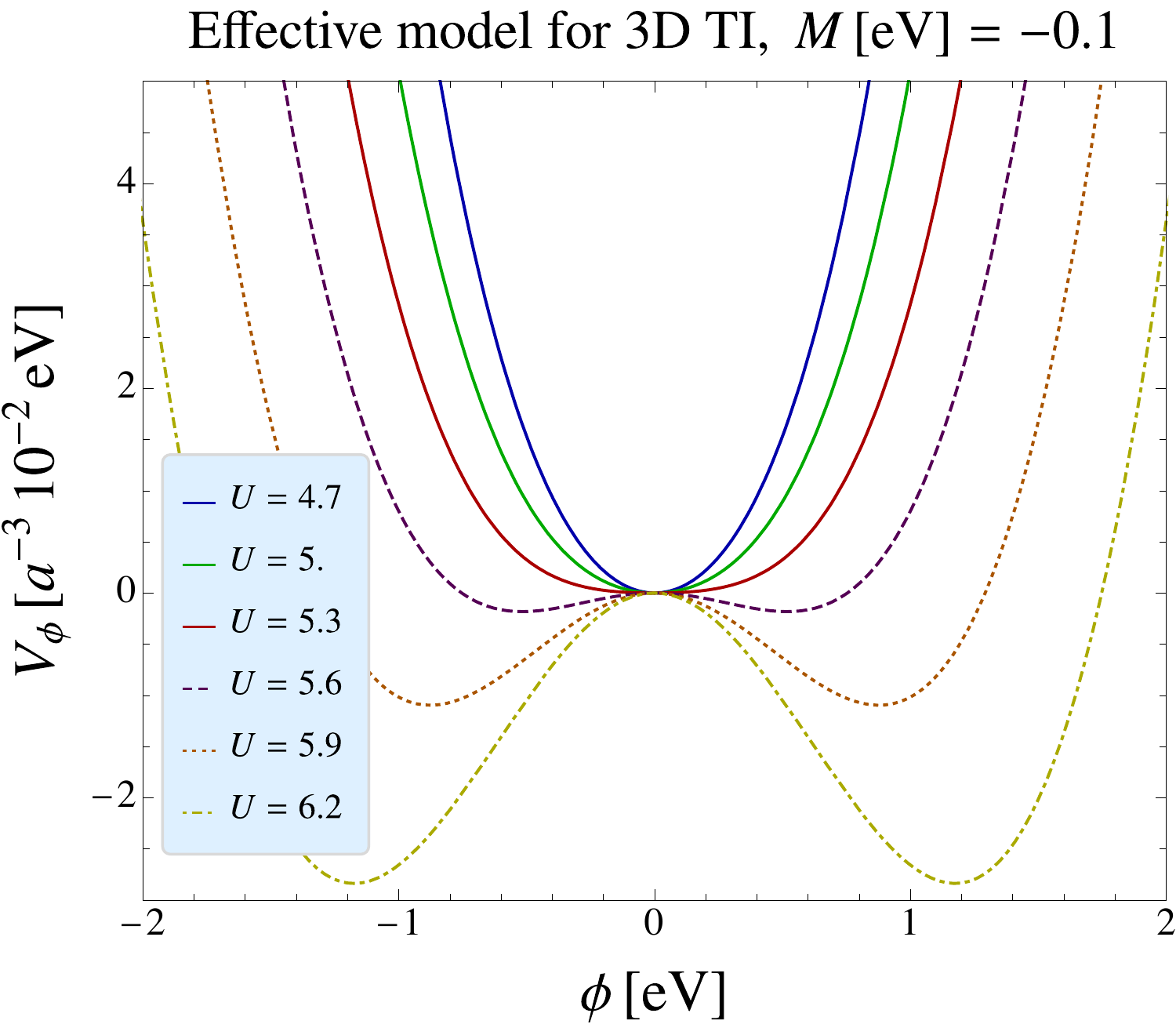}
    \includegraphics[scale=0.38]{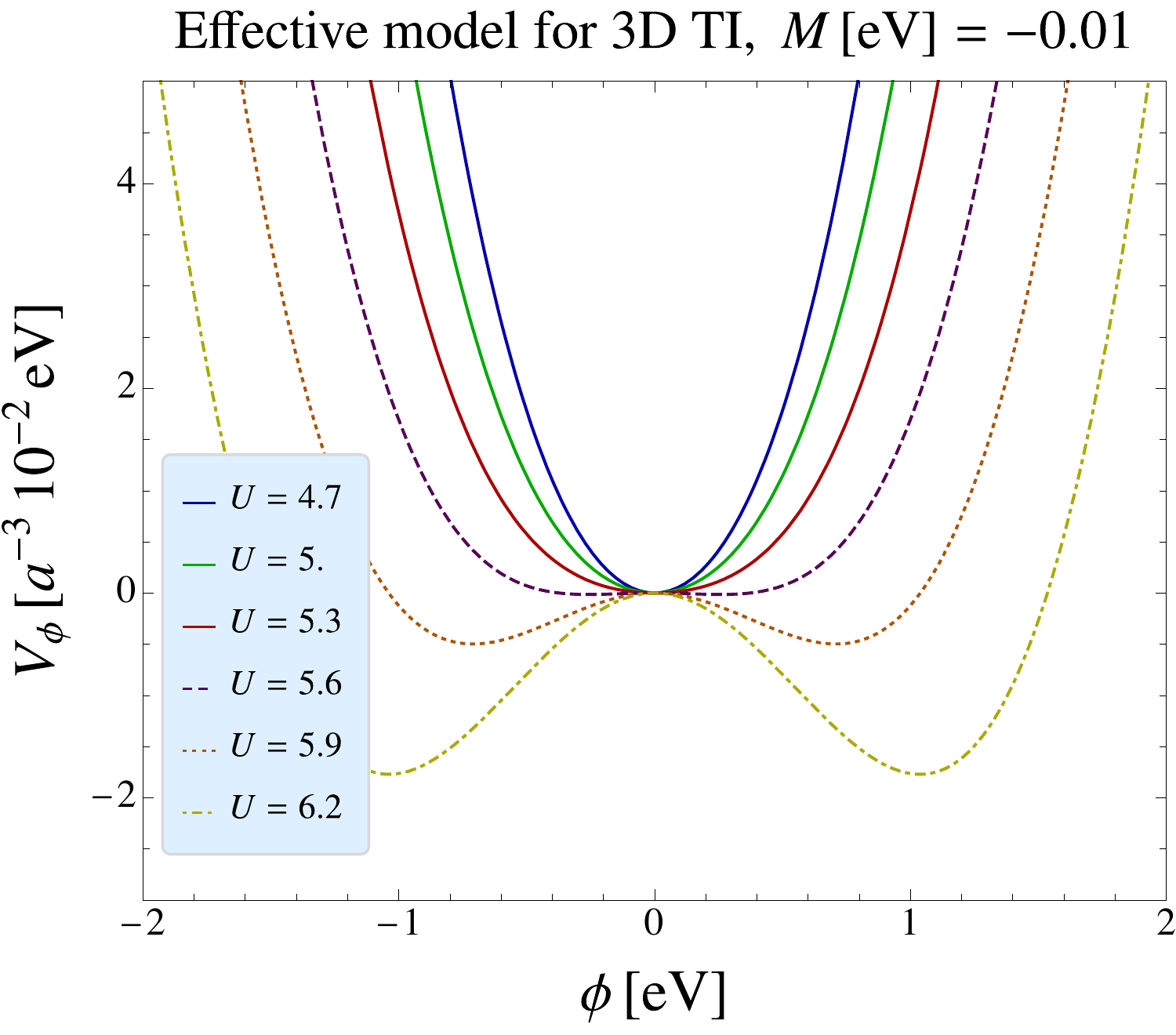}
    \includegraphics[scale=0.38]{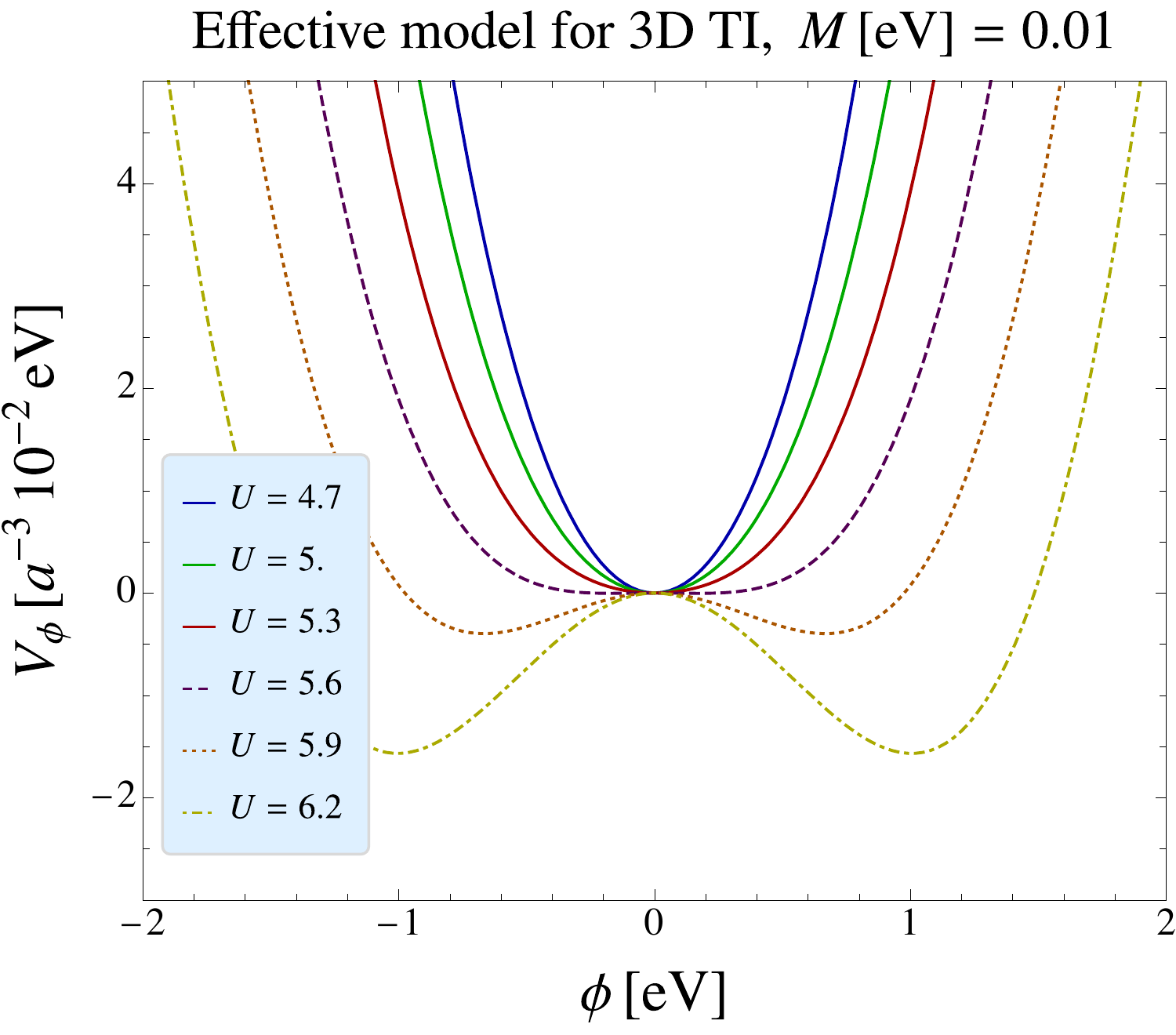}
    \includegraphics[scale=0.38]{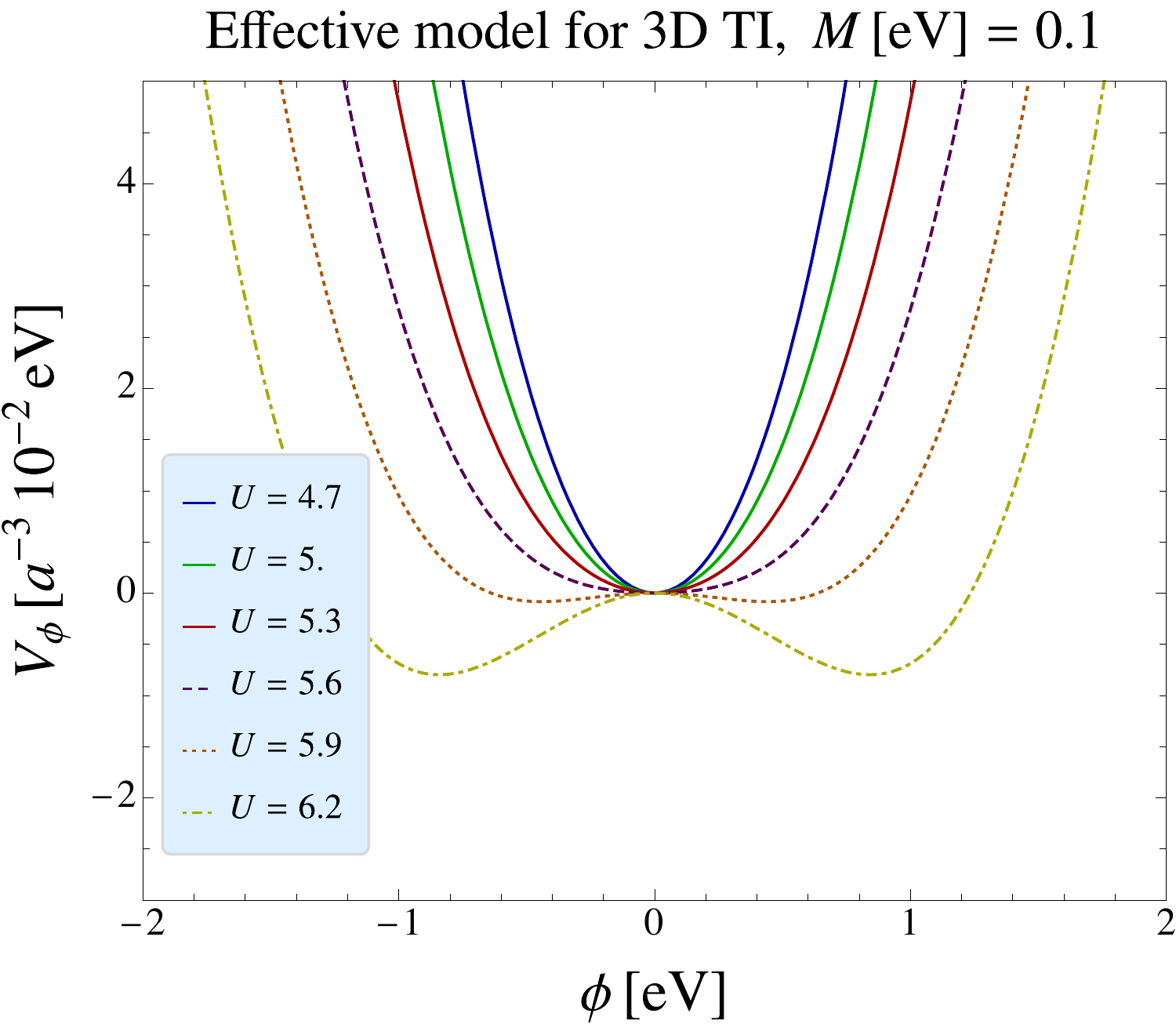}
    \includegraphics[scale=0.38]{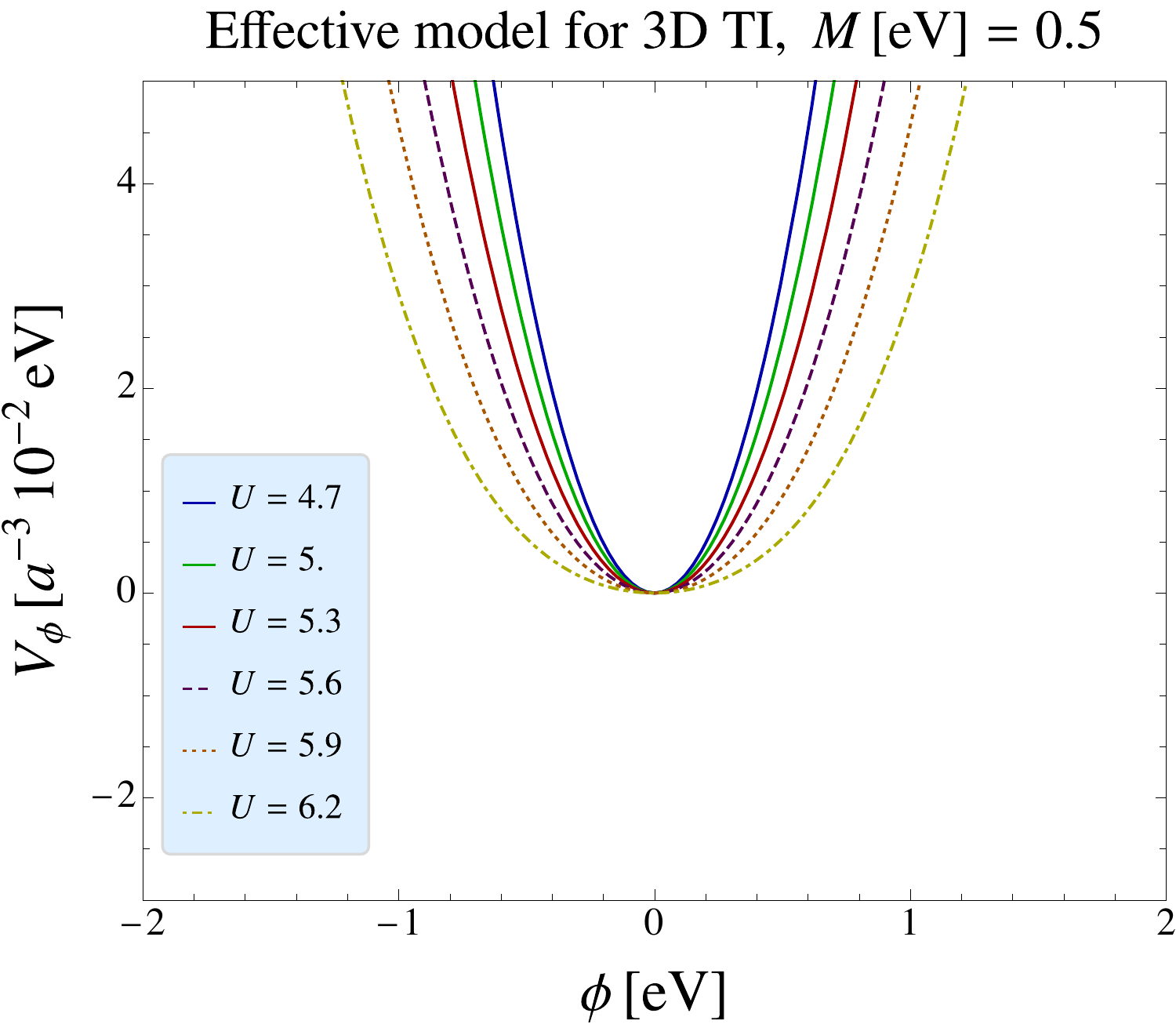}
   \end{center}
  \caption{$V_\phi$ as function of $\phi$ for various values of $M$ in
    the effective model for 3D topological insulators.  $A_1$, $A_2$,
    $B_1$, $B_2$, and $M$ are taken the same as
    Fig.\,\ref{fig:theta_phi_compare}. }
  \label{fig:JVphi_Li}
\end{figure}

\begin{figure}[t]
  \begin{center}
    \includegraphics[scale=0.38]{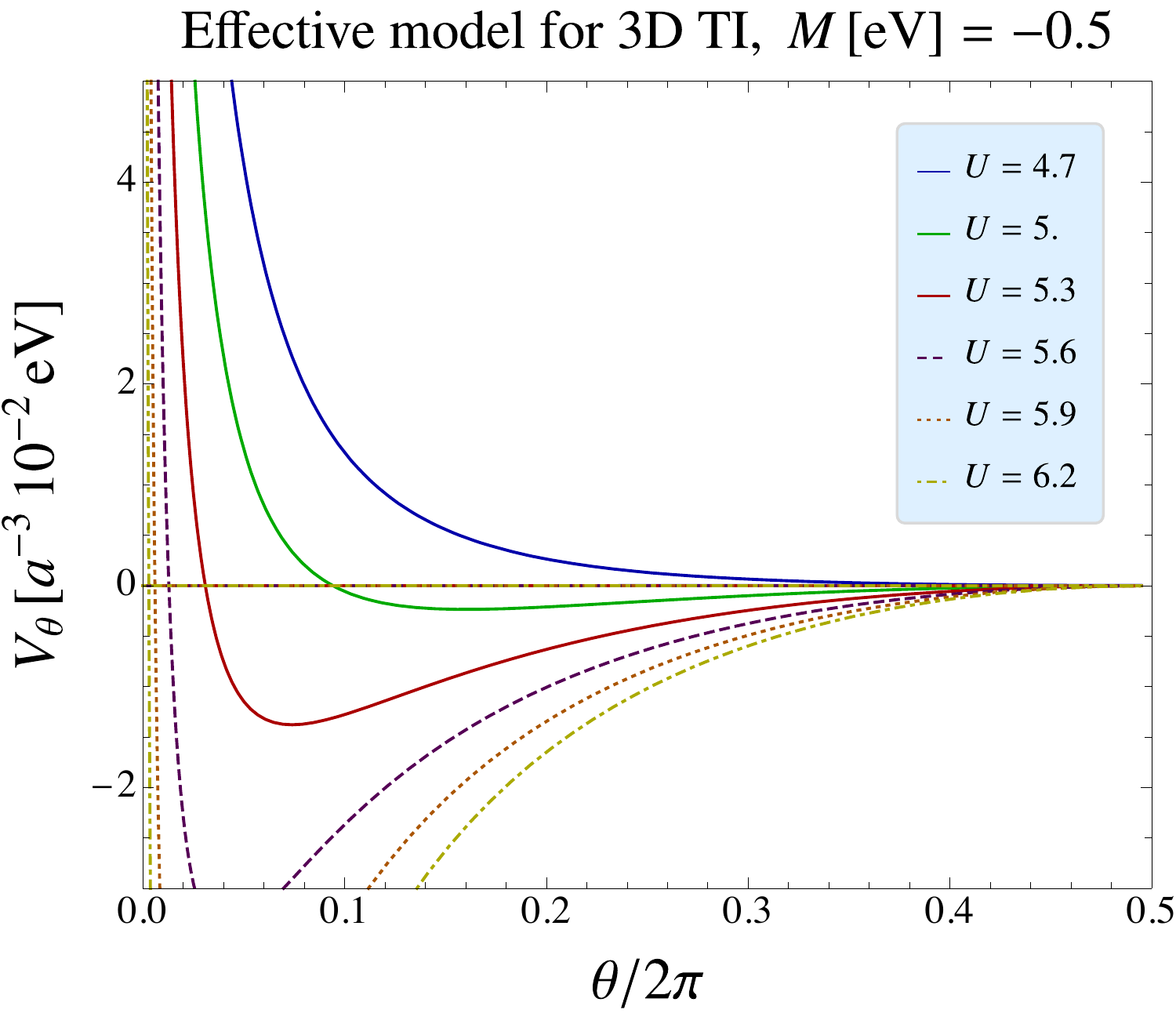}
    \includegraphics[scale=0.38]{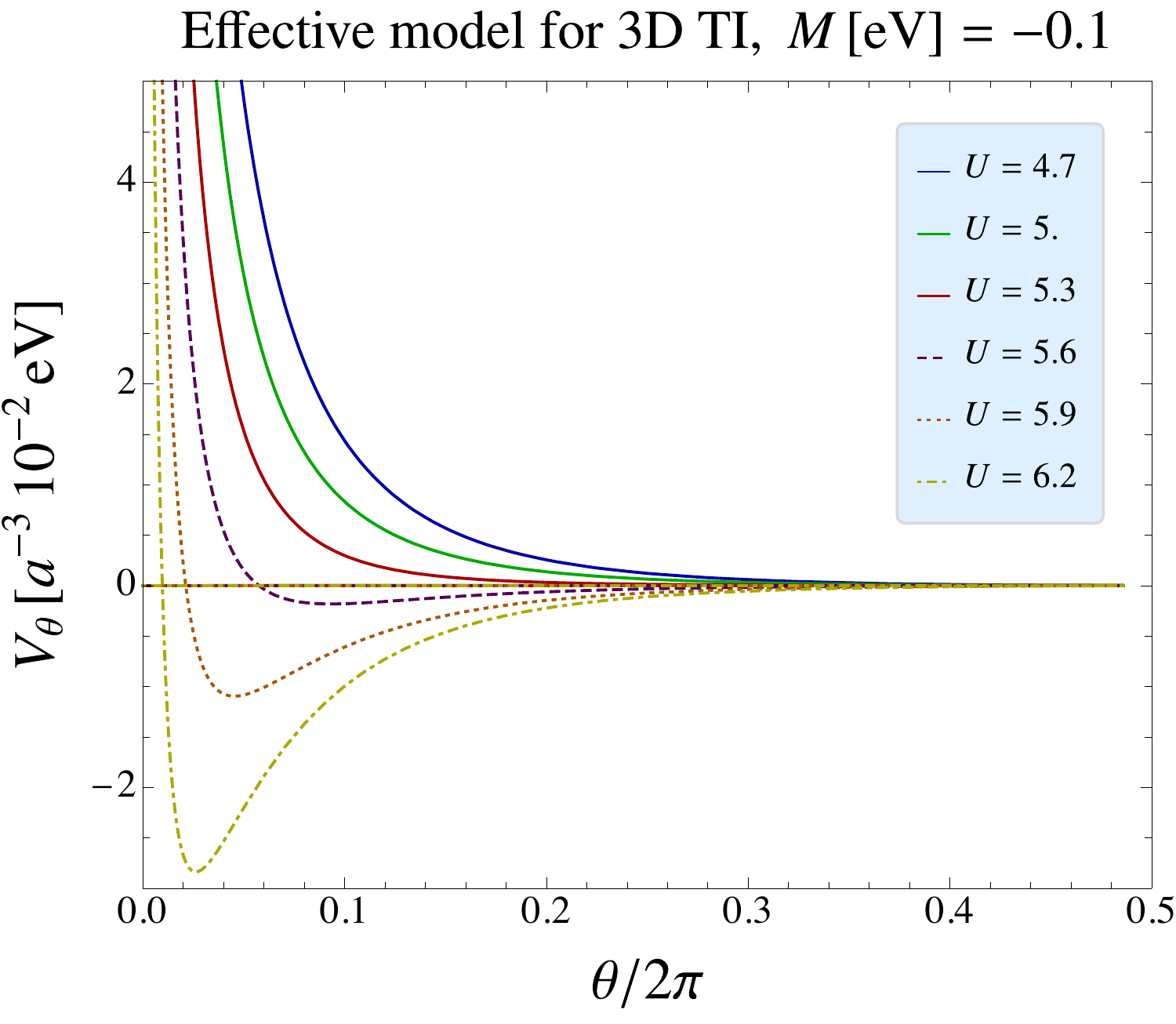}
    \includegraphics[scale=0.38]{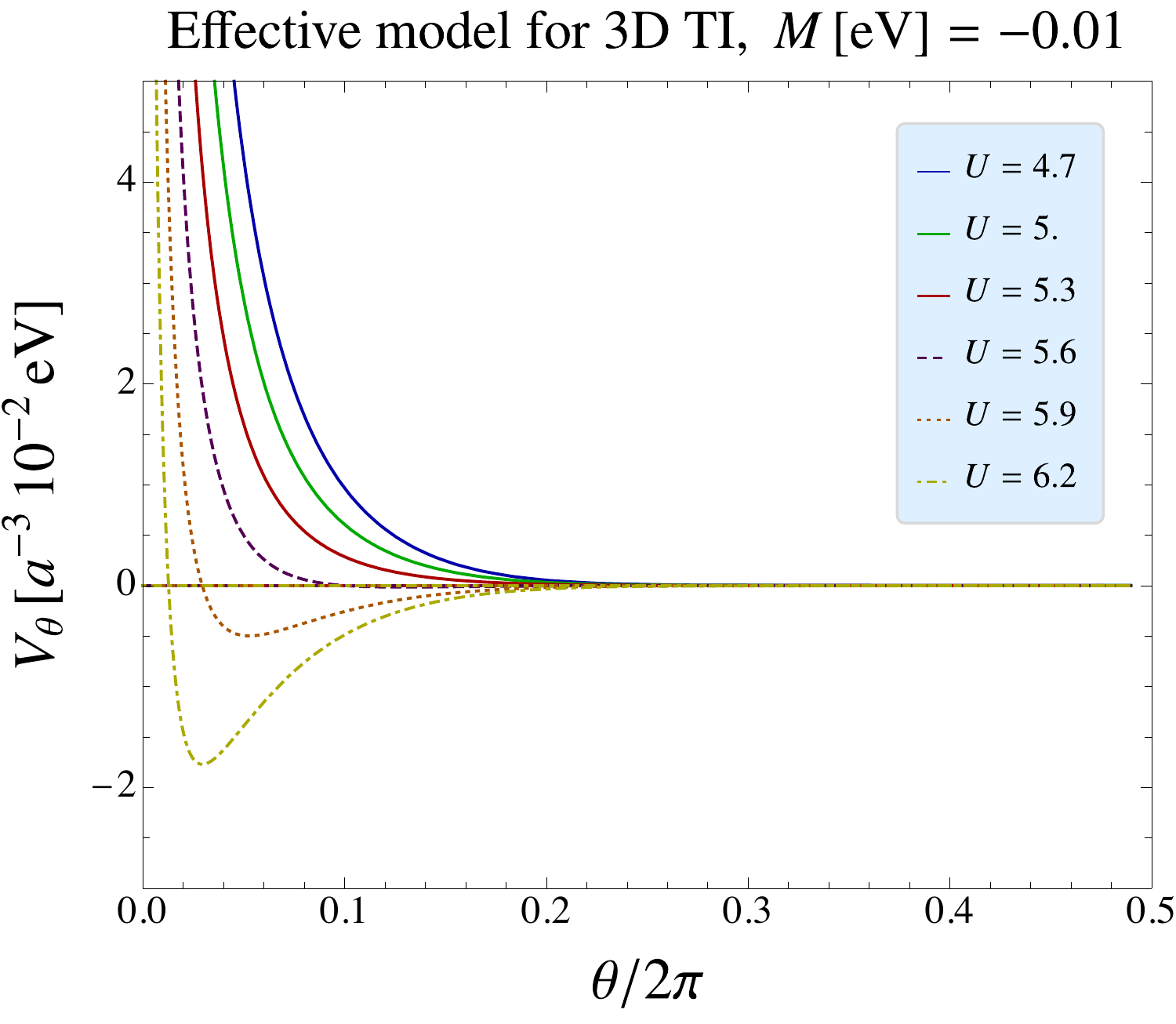}
    \includegraphics[scale=0.38]{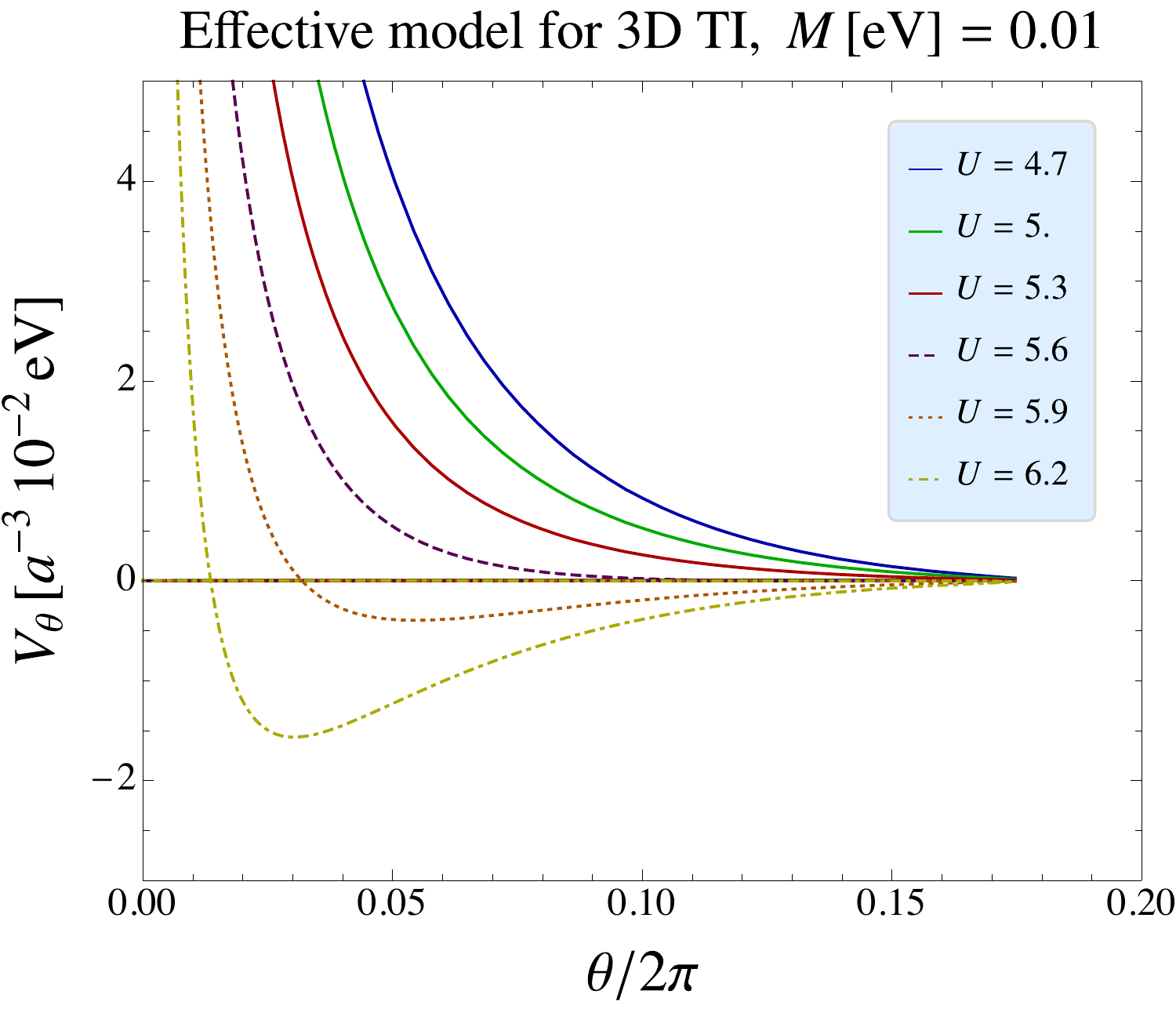}
    \includegraphics[scale=0.38]{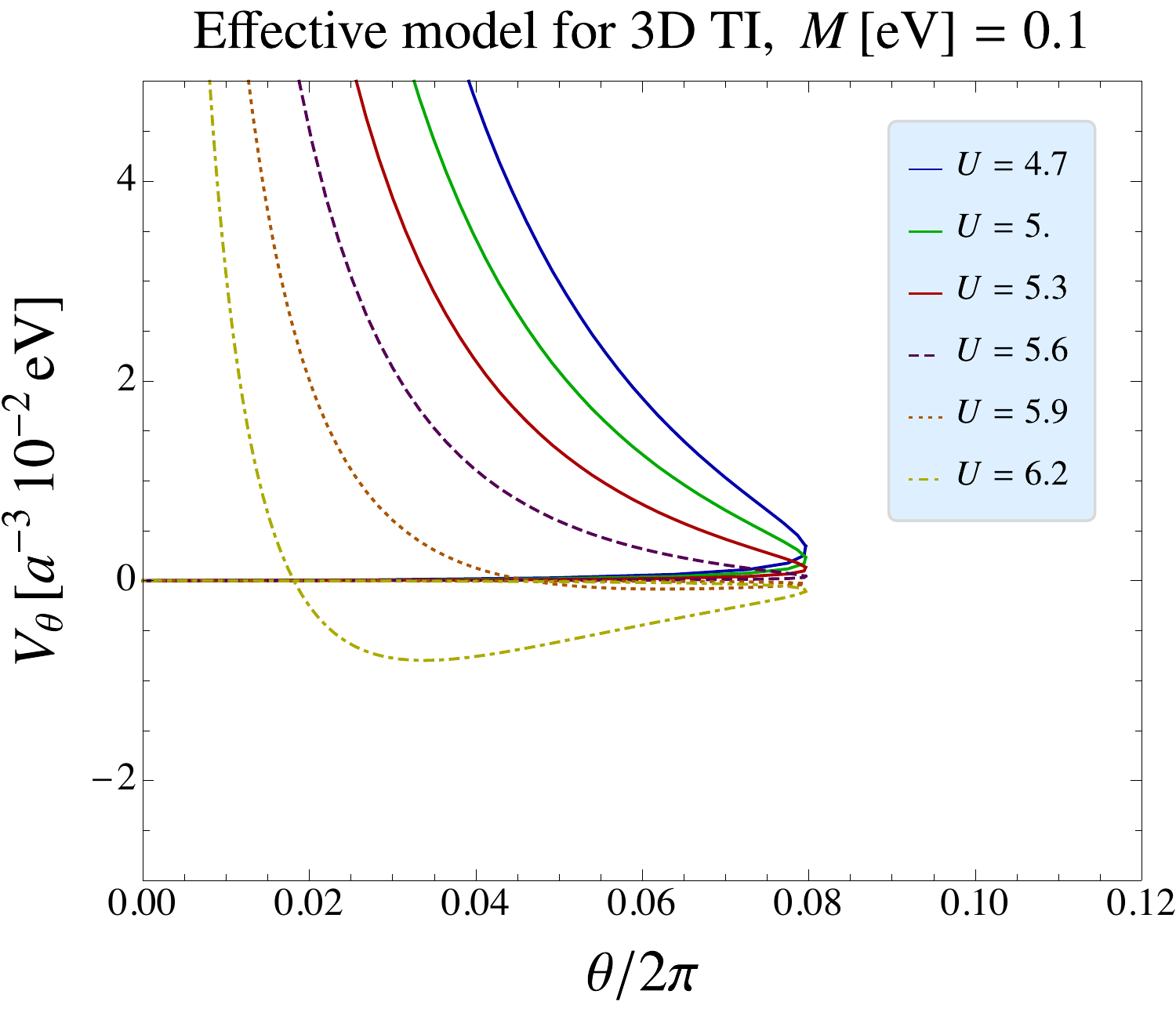}
    \includegraphics[scale=0.38]{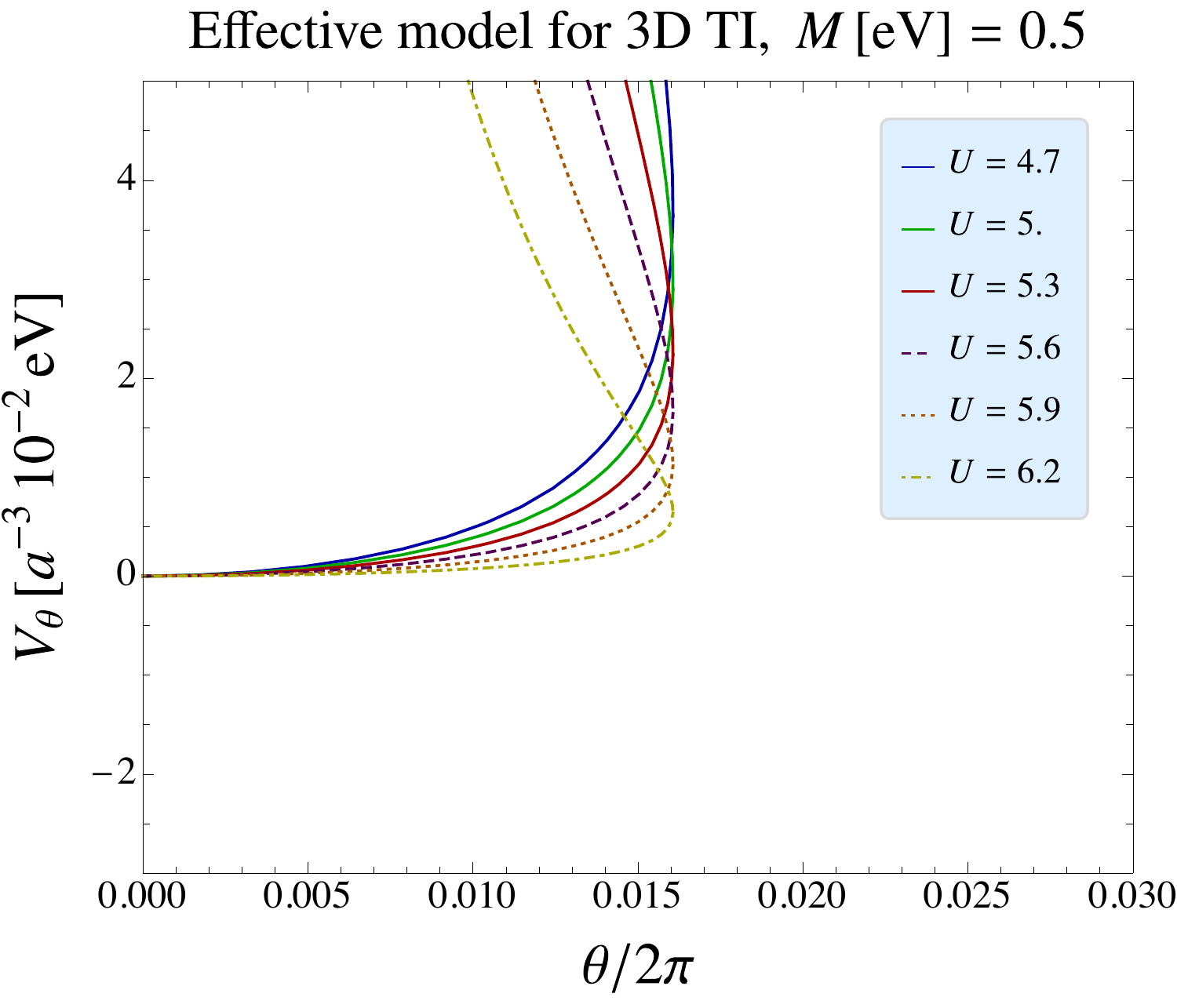}
  \end{center}
  \caption{$V_\theta$ as function of $\theta/(2\pi)$ in the effective model
    for 3D topological insulators.  Parameters are taken the same as
    Fig.\,\ref{fig:JVphi_Li}. }
  \label{fig:JVtheta_Li}
\end{figure}

\begin{figure}[t]
  \begin{center}
    \includegraphics[scale=0.45]{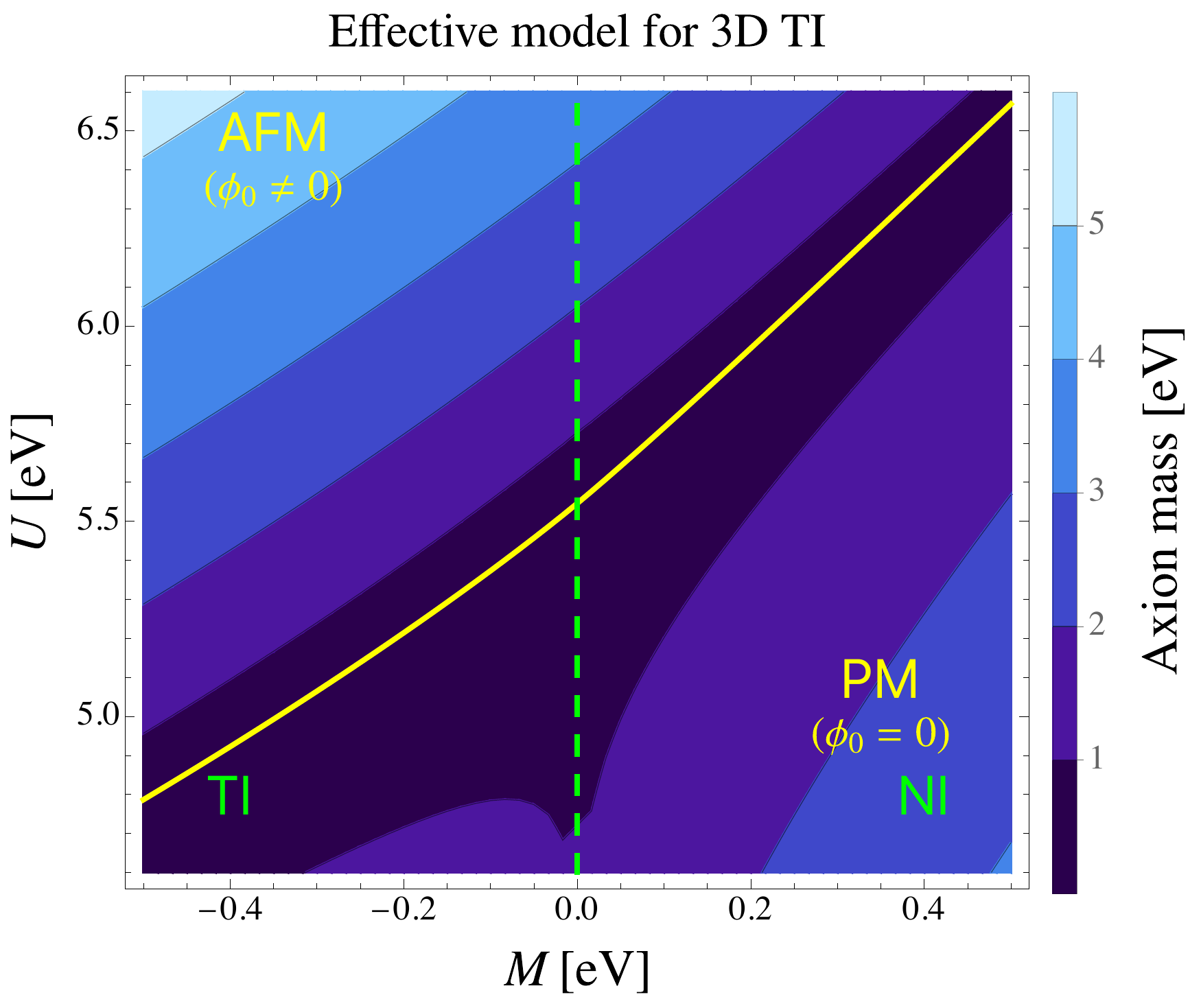}
      \end{center}
  \caption{Axion mass on ($U$,\,$M$) plane calculated in the
    effective model for 3D topological insulators.
    $A_1/a^2=A_2/a^2=1$~eV and $B_1/a^2=B_2/a^2=-0.5$~eV are
    taken. Solid (yellow) line shows the critical points; upper and
    lower regions separated by the solid line correspond to
    antiferromagnetic order (``AFM'') and paramagnetic order (``PM''),
    respectively. Dotted (green) line is boundary of topological
    insulator phase (``TI'') and normal insulator phase (``NI''). }
  \label{fig:ma_Li}
\end{figure}

\section{Origin of $d^5\Gamma^5$ term}
\label{sec:m5term}
\setcounter{equation}{0} 

In this section, we discuss a possible origin of $\delta H$ term given
in Eq.\,\eqref{eq:deltaH}. For pedagogical purpose, we provide three
types of derivations; derivations in the continuum space (Appendix
\ref{sec:continuum_case}), in a discrete space (Appendix
\ref{sec:discrete_case}) and by the mean field approximation (Appendix
\ref{sec:MFA}).

\subsection{In continuum space}
\label{sec:continuum_case}
\setcounter{equation}{0}

To begin with, we consider one of sublattices since it is enough to
get the basic idea of emerging a scalar degree of freedom that
corresponds to $m_5\,(=\phi)$. The interaction term to cause the AFM order
is given by following Hamiltonian:\footnote{This term corresponds to
the continuum limit of Eq.\,\eqref{eq:H_I_discrete} by making a
replacement, $\frac{1}{N}\sum_i\to\frac{1}{V}\int d^3x$ and
$n_{i\sigma}\to\frac{V}{N}n_\sigma(\vb*{x})$.}
\begin{align}
  {\cal H}_{\rm int} =
  \frac{U V}{N}\int \dd[3]{x}n_\uparrow(\vb*{x})n_\downarrow(\vb*{x})\,,
\end{align}
where arrow stands for the spin, $V$ is the volume of the material,
$N$ is the number of site, and
\begin{align}
  n_\sigma(\vb*{x})=\psi^\dagger_\sigma \psi_\sigma
  ~~~~~~ (\sigma =\, \uparrow\,,\downarrow)~\,.
\end{align}
It is convenient to define the following quantities:
\begin{align}
  n_\pm(\vb*{x}) \equiv
  n_\uparrow(\vb*{x})\pm n_\downarrow(\vb*{x})\,,
\end{align}
and take
\begin{align}
  \expval{n_+(\vb*{x})}= {\rm const.}\,,
  \label{eq:n+=const}
\end{align}
since the total number of electrons are unchanged. (This corresponds
to Eq.\,\eqref{eq:totalnumberdensity_MFA} in the mean field
approximation.) Therefore, only $n_-$ is a dynamical degree of freedom,
and ${\cal H}_{\rm int}$ becomes\footnote{Strictly speaking, there is
always a fluctuation around $\expval{n_+}$. However, it is not one we
are interested in. See also the discussion when two sublattices are
introduced.}
\begin{align}
  {\cal H}_{\rm int}=-\frac{UV}{4N}\int \dd[3]{x} n_-^2(\vb*{x})
  + {\rm const.}
\end{align}

For later discussion we define the Fourier expansion of $\psi_\sigma$
and $n_\sigma$ as
\begin{align}
  \psi_\sigma(\vb*{x})&=\frac{1}{\sqrt{V}}\sum_{\vb*{k}}
  c_{\vb*{k}\sigma}e^{i\vb*{k}\vdot \vb*{x}}\,,
  \\
  c_{\vb*{k}\sigma}&=\frac{1}{\sqrt{V}}\int \dd[3]{x}
  \psi_{\sigma}(\vb*{x})e^{-i\vb*{k}\vdot \vb*{x}}\,,
  \\
  n_\sigma(\vb*{x}) &=
  \frac{1}{V}\sum_{\vb*{k}}
  \tilde{n}_\sigma(\vb*{k})e^{i\vb*{k}\vdot \vb*{x}}\,,
  \\
  \tilde{n}_\sigma(\vb*{k})&=\int \dd[3]{x} n_\sigma (\vb*{x})
  e^{-i \vb*{k}\cdot \vb*{x}}
=\sum_{\vb*{q}}c^\dagger_{\vb*{q}\sigma}c_{\vb*{q}+\vb*{k}\sigma}
  \,.
\end{align}
Consequently, ${\cal H}_{\rm int}$ in the momentum space is given by
\begin{align}
  {\cal H}_{\rm int} =
   \frac{U}{N}\sum_{\vb*{k}}
   \tilde{n}_\uparrow(\vb*{k}) \tilde{n}_\downarrow(-\vb*{k})
   =
    -\frac{U}{4N}\sum_{\vb*{k}}
    \tilde{n}_-(\vb*{k}) \tilde{n}_-(-\vb*{k})
    +{\rm const.}
\end{align}

In terms of Euclidean time $\tau=it$, action of the interaction term is
written by 
\begin{align}
  -S_{{\rm int}\,E}\equiv iS_{\rm int}
  \equiv-
  \int \dd{\tau} {\cal H}_{\rm int}\,.
\end{align}
Now we execute the Hubbard-Stratonovich transformation. It is also
used in Ref.\,\cite{Shiozaki:2013wda}, which studies topological
superconductors and superfluids. (See Ref.\,\cite{Roy:2015xua} for the
renormalization group approach.) It is effectively the same as the
inverse procedure of integrating out the heavy scalar particles to
give the effective Lagrangian. (See
Appendix~\ref{sec:Lagrangian_in_DiracRep} for details.) Introducing
$\tilde{\phi}(\vb*{k})$ and using the following identity,
\begin{align}
  &
  -\Bigl\{\frac{N}{U}\tilde{\phi}(\vb*{k})\tilde{\phi}(-\vb*{k})
  +
  \frac{1}{2}
  \left[\tilde{\phi}(\vb*{k})\tilde{n}_-(-\vb*{k})
    +\tilde{\phi}_I(-\vb*{k})\tilde{n}_-(\vb*{k})
    \right]\Bigr\}
  \nonumber \\
  &=
  -\frac{N}{U}
  \left[\tilde{\phi}(\vb*{k})+\frac{U}{2N}\tilde{n}_-(\vb*{k})\right]
  \left[\tilde{\phi}(-\vb*{k})+\frac{U}{2N}\tilde{n}_-(-\vb*{k})\right]
  +
  \frac{U}{4N}
  \tilde{n}_-(\vb*{k})\tilde{n}_-(-\vb*{k})\,,
  \label{eq:ForHStr}
\end{align}
the Euclidean action gives
\begin{align}
  &\exp\left[-\int \dd{\tau} {\cal H}_{\rm int}
    \right]
  \nonumber \\
  &=\int {\cal D}\tilde{\phi}
  \exp\left\{-\int \dd{\tau} \left\{
    \sum_{\vb*{k}}
    \frac{N}{U}
    \tilde{\phi}(\vb*{k})\tilde{\phi}(-\vb*{k})
  +
  \frac{1}{2}\sum_{\vb*{k}}
  \left[\tilde{\phi}(\vb*{k})\tilde{n}_-(-\vb*{k})
    +\tilde{\phi}(-\vb*{k})\tilde{n}_-(\vb*{k})
    \right]
  \right\}
  \right\}
  \nonumber \\
  &=\int {\cal D}\phi
  \exp\left\{
  -\int \dd{\tau} \dd[3]{x}
  \left[\frac{N}{VU}\phi^2(\vb*{x})+\phi(\vb*{x})
    n_-(\vb*{x})
    \right]
  \right\}\,.
  \label{eq:S_int}
\end{align}
Here we have omitted irrelevant constant term in the last step and we
have used
\begin{align}
  \tilde{\phi}(\vb*{k})=
  \frac{1}{V}\int \dd[3]{x} \phi(x) e^{-i \vb*{k}\cdot \vb*{x}}\,.
\end{align}
The second term on the rhs of Eq.\,\eqref{eq:S_int} is given in the
momentum space as
\begin{align}
  \int \dd[3]{x} \phi(\vb*{x})n_-(\vb*{x})
  &=\sum_{\vb*{q},\vb*{k}}
  (c^\dagger_{\vb*{k}\uparrow}c_{\vb*{k}+\vb*{q}\uparrow}
  -c^\dagger_{\vb*{k}\downarrow}c_{\vb*{k}+\vb*{q}\downarrow})
  \tilde{\phi}(-\vb*{q})
  \nonumber \\
  &=\sum_{\vb*{k}}
   (c^\dagger_{\vb*{k}\uparrow}c_{\vb*{k}\uparrow}
  -c^\dagger_{\vb*{k}\downarrow}c_{\vb*{k}\downarrow})
  \tilde{\phi}(0)+{\cal O}(\vb*{q})
  \nonumber \\
  &=\sum_{\vb*{k},\alpha,\beta}
  c^\dagger_{\vb*{k}\alpha}
  (\tilde{\phi}(0) \sigma^z)_{\alpha\beta}
  c_{\vb*{k}\beta}
  +{\cal O}(\vb*{q})\,.
  \label{eq:int_term_k}
\end{align}

Now we apply the above result to two sublattices, A and B. The interaction
term  to start with is
\begin{align}
   {\cal H}_{\rm int}^{{\rm A}+{\rm B}} =
   \frac{U V}{N}\int \dd[3]{x}
   (
   n_{{\rm A}\uparrow}(\vb*{x})n_{{\rm A}\downarrow}(\vb*{x})
   +n_{{\rm B}\uparrow}(\vb*{x})n_{{\rm B}\downarrow}(\vb*{x})
   )
   \,,
\end{align}
where $N$ is the number of the sublattice A (B). As in the previous
discussion, we define $n_{{\rm I}\pm}(\vb*{x})$ (I=A, B),
\begin{align}
  n_{{\rm I}\pm}(\vb*{x})\equiv
  n_{{\rm I}\uparrow}(\vb*{x})\pm n_{{\rm I}\downarrow}(\vb*{x})\,.
\end{align}
Since we are interested in the AFM order, $n_{{\rm A}-}$ and $n_{{\rm
    B}-}$ are not independent and related by\footnote{As discussed
below Eq.\,\eqref{eq:n+=const}, there are fluctuations around
$\expval{n_{{\rm A}+}}$ and $\expval{n_{{\rm B}+}}$. In addition, if
there is no the AFM order, {\it i.e.}, the PM order, then
$\expval{n_{{\rm A}-}}$ and $\expval{n_{{\rm B}-}}$ should be treated
as independent degrees of freedom. }
\begin{align}
  n_{{\rm A}-} +n_{{\rm B}-}\approx
  \expval{n_{{\rm A}-} +n_{{\rm B}-}}=0\,.
  \label{eq:AssumingAFM}
\end{align}
Therefore, we obtain
\begin{align}
  &\exp\left[-\int \dd{\tau} {\cal H}_{\rm int}^{{\rm A}+{\rm B}}
    \right]
  \nonumber \\
  &=\int {\cal D}\tilde{\phi}
  \exp\left\{-\int \dd{\tau} \left\{
    \sum_{\vb*{k}}
    \frac{2N}{U}
    \tilde{\phi}(\vb*{k})\tilde{\phi}(-\vb*{k})
  +
  \frac{1}{2}\sum_{\vb*{k}}
  \left[\tilde{\phi}(\vb*{k})
    (\tilde{n}_{{\rm A}-}(-\vb*{k})
    -\tilde{n}_{{\rm B}-}(-\vb*{k}))
    +\tilde{\phi}(-\vb*{k})
    (\tilde{n}_{{\rm A}-}(\vb*{k})
    -\tilde{n}_{{\rm B}-}(\vb*{k}))
    \right]
  \right\}
  \right\}
  \nonumber \\
  &=\int {\cal D}\phi
  \exp\left\{
  -\int \dd{\tau} \dd[3]{x}
  \left[\frac{2N}{VU}\phi^2(\vb*{x})+\phi(\vb*{x})
    (n_{{\rm A}-}(\vb*{x})-n_{{\rm B}-}(\vb*{x}))
    \right]
  \right\}\,.
  \label{eq:H_int^A+B}
\end{align}
and
\begin{align}
  \int \dd[3]{x} \phi(\vb*{x})
  (n_{{\rm A}-}(\vb*{x})-n_{{\rm B}-}(\vb*{x}))
   =\sum_{\vb*{k}}\tilde{\phi}(0) 
   \left\{
     c^\dagger_{{\rm A}\vb*{k}}
     \sigma^z
     c_{{\rm A}\vb*{k}}
     -
     c^\dagger_{{\rm B}\vb*{k}}
     \sigma^z
     c_{{\rm B}\vb*{k}}
     \right\}
  +{\cal O}(\vb*{q})\,.
  \label{eq:int_term_k_tot}
\end{align}
Here the summation over spin indices are implicit.  This term
corresponds to $d^5\Gamma^{5\prime}\,(=\phi \Gamma^{5\prime})$ term in
Eq.\,\eqref{eq:Hk} after changing the basis from ($\ket{
  P1_z^+,\sigma}$, $\ket{P2_z^-,\sigma}$) to $(\ket{{\rm
    A},\sigma},\ket{{\rm B},\sigma})$ where $\sigma
=\,\uparrow,\downarrow$.  In this basis, ${\Gamma}^{5\prime} = {\rm
  diag}\{1,-1,-1,1\}$ and Eq.\,\eqref{eq:H_int^A+B} becomes
\begin{align}
  &\int {\cal D}\phi
  \exp\left\{-
  \int \dd{\tau} \dd[3]{x} \left[ M^2 \phi^2(\vb*{x}) -
    \phi(\vb*{x})
    (n_{{\rm A}-}(\vb*{x})-n_{{\rm B}-}(\vb*{x}))
     \right]+\cdots
  \right\}
  \nonumber \\
  &=\int {\cal D}\tilde{\phi}
  \exp\left\{-\int \dd{\tau}
  \sum_{\vb*{k}}\left[VM^2\tilde{\phi}(\vb*{k})\tilde{\phi}(-\vb*{k})
  -\vb*{c}^\dagger_{\vb*{k}}
  \Gamma^{5\prime}\tilde{\phi}(0)\vb*{c}_{\vb*{k}} +\cdots
  \right]
  \right\}\,,
  \label{eq:H_A+H_B_x}
\end{align}
where $\vb*{c}_{\vb*{k}}=(c_{{\rm A}\vb*{k}\uparrow},\,c_{{\rm A}\vb*{k}\downarrow
},\,c_{{\rm B}\vb*{k}\uparrow },\,c_{{\rm B}\vb*{k}\downarrow })$ and\footnote{We
use $(V/N)^{-1}=\int \frac{d^3q}{(2\pi)^3}$ where the integral is
defined in the first Brillouin zone.}
\begin{align}
  M^2\equiv \frac{2}{(V/N)U} =\left(\int
  \frac{d^3q}{(2\pi)^3}\right)\frac{2}{U} \,.
  \label{eq:M^2_derive}
\end{align}

\subsection{In discrete space}
\label{sec:discrete_case}

As in the previous subsection, we first consider one of the sublattices
and extend the result to the other later.  In the discrete space, we
use annihilation operator $c_{i\sigma}$ where $i$ is a label of site
instead of the wavefunction $\psi_\sigma(\vb*{x})$. The interaction
Hamiltonian is
\begin{align}
  {\cal H}_{\rm int} =
  U\sum_i n_{i\uparrow}n_{i\downarrow}\,,
  \label{eq:H_I_discrete}
\end{align}
where
\begin{align}
  n_{i\sigma} = c_{i\sigma}^\dagger c_{i\sigma}\,.
\end{align}
While defining $n_{i\pm}$ as
\begin{align}
  n_{i\pm}\equiv n_{i\uparrow}\pm n_{i\downarrow}\,,
\end{align}
only $n_{i-}$ is dynamical since $n_{i+}=$const. as in
Eq.\,\eqref{eq:n+=const}. Consequently, the interaction Hamiltonian is
given by
\begin{align}
  {\cal H}_{\rm int} =
  \frac{U}{4}\sum_i n_{i-}^2+{\rm const.}\,.
\end{align}
It is useful to give Fourier transformation of $c_{i\sigma}$ and
$n_{i\sigma}$\,,
\begin{align}
  c_{i\sigma} &=\frac{1}{\sqrt{N}}
  \sum_{\vb*{k}}c_{\vb*{k}\sigma}e^{i\vb*{k}\vdot \vb*{x}}\,,
  \\
  c_{\vb*{k}\sigma} &=\frac{1}{\sqrt{N}}
  \sum_{i}c_{i\sigma}e^{-i\vb*{k}\vdot \vb*{x}}\,,
  \\
  n_{i\sigma}&=\frac{1}{N}\sum_{\vb*{k}}
  \tilde{n}_{\vb*{k}}e^{i\vb*{k}\vdot \vb*{x}}\,,
  \\
  \tilde{n}_{\vb*{k}\sigma}
  &=\sum_in_{i\sigma}e^{-i\vb*{k}\vdot \vb*{x}}
  =\sum_{\vb*{q}}c^\dagger_{\vb*{q}\sigma}c_{\vb*{q}+\vb*{k}\sigma}
  \,,
\end{align}
which leads to the Hamiltonian in the momentum space,
\begin{align}
  {\cal H}_{\rm int}
  = \frac{U}{N}\sum_{\vb*{k}}
  \tilde{n}_{\vb*{k}\uparrow}\tilde{n}_{-\vb*{k}\downarrow}
  =-\frac{U}{4N}\sum_{\vb*{k}}
  \tilde{n}_{\vb*{k}-}\tilde{n}_{-\vb*{k}-}
  +{\rm const.}\,
\end{align}
As in the continuum case, the Hubbard-Stratonovich transformation is done
by using Eq.\eqref{eq:ForHStr}. Then the Euclidean action of the
interaction term is given by
\begin{align}
  &\exp\left[-\int d\tau {\cal H}_{\rm int}
    \right]
  \nonumber \\
  &=\int {\cal D}\tilde{\phi}
  \exp\left\{-\int \dd{\tau} \left\{
  \sum_{\vb*{k}}
  \frac{N}{U}
  \tilde{\phi}_{\vb*{k}}\tilde{\phi}_{-\vb*{k}}
  +
  \sum_{\vb*{k}}\frac{1}{2}
  \left[\tilde{\phi}_{\vb*{k}}\tilde{n}_{-\vb*{k}-}
    +\tilde{\phi}_{-\vb*{k}}\tilde{n}_{\vb*{k}-}
    \right]
  \right\}
  \right\}
  \nonumber \\
  &=\int {\cal D}\tilde{\phi}
  \exp\left\{
  -\int \dd{\tau} \sum_i
  \left[\frac{1}{U}\phi_i^2+\phi_{i}
    n_{i-}
    \right]
  \right\}\,.
  \label{eq:S_int_discrete}
\end{align}
Here we have defined the Fourier expansion of $\phi_{i}$ as
\begin{align}
  \phi_{i}&=\sum_{\vb*{k}}\tilde{\phi}_{\vb*{k}}e^{i\vb*{k}\vdot \vb*{x}}\,,
  \\
  \tilde{\phi}_{\vb*{k}}
  &=\frac{1}{N}\sum_i\phi_{i}e^{-i\vb*{k}\vdot \vb*{x}}\,,
\end{align}
and the second term on the rhs of Eq.\,\eqref{eq:S_int_discrete} is
written in the momentum space as
\begin{align}
  \sum_i \phi_{i} n_{i-}
  &=\sum_{\vb*{q},\vb*{k}}
  (c^\dagger_{\vb*{k}\uparrow}c_{\vb*{k}+\vb*{q}\uparrow}
  -c^\dagger_{\vb*{k}\downarrow}c_{\vb*{k}+\vb*{q}\downarrow})
  \tilde{\phi}_{-\vb*{q}}
  \nonumber \\
  &=\sum_{\vb*{k}}
  (c^\dagger_{\vb*{k}\uparrow}c_{\vb*{k}\uparrow}
  -c^\dagger_{\vb*{k}\downarrow}c_{\vb*{k}\downarrow})
  \tilde{\phi}_{\vb{0}}+{\cal O}(\vb*{q})
  \nonumber \\
  &=\sum_{\vb*{k},\alpha,\beta}
  c^\dagger_{\vb*{k}\alpha}
  (\tilde{\phi}_{\vb{0}} \sigma^z)_{\alpha\beta}
  c_{\vb*{k}\beta}
  +{\cal O}(\vb*{q})\,.
  \label{eq:int_term_k_discrete}
\end{align}

Extension to the other sublattice is trivial. As in the previous subsection,
let us call the two sublattice as A and B. Then the interaction
Hamiltonian is
\begin{align}
  {\cal H}_{\rm int}^{\rm A+B}=
  U\sum_i (
  n_{{\rm A}i\uparrow}n_{{\rm A}i\downarrow}
  +n_{{\rm B}i\uparrow}n_{{\rm B}i\downarrow})\,.
\end{align}
Repeating the discussion so far, the Euclidean action is obtained as
\begin{align}
    &\exp\left[-\int \dd{\tau} {\cal H}_{\rm int}^{{\rm A}+{\rm B}}
    \right]
  \nonumber \\
  &=\int {\cal D}\tilde{\phi}
  \exp\left\{-\int \dd{\tau} \left\{
    \sum_{\vb*{k}}
    \frac{2N}{U}
    \tilde{\phi}_{\vb*{k}}\tilde{\phi}_{-\vb*{k}}
  +
  \frac{1}{2}\sum_{\vb*{k}}
  \left[\tilde{\phi}_{\vb*{k}}
    (\tilde{n}_{{\rm A}-\vb*{k}-}
    -\tilde{n}_{{\rm B}-\vb*{k}-})
    +\tilde{\phi}_{-\vb*{k}}
    (\tilde{n}_{{\rm A}\vb*{k}-}
    -\tilde{n}_{{\rm B}\vb*{k}-})
    \right]
  \right\}
  \right\}
  \nonumber \\
  &=\int {\cal D}\phi
  \exp\left\{
  -\int \sum_i
  \left[\frac{2}{U}\phi^2_i+\phi_i
    (n_{i{\rm A}-}-n_{i{\rm B}-})
    \right]
  \right\}\,.
  \label{eq:H_int^A+B_discrete}
\end{align}
The interaction part in momentum space is
\begin{align}
  \sum_i \phi_{i}
    (n_{i\rm{A}-}-n_{i\rm{B}-})
  =
  \sum_{\vb*{k}}
  \tilde{\phi}_{\vb{0}}
  \left\{
     c^\dagger_{{\rm A}\vb*{k}}
      \sigma^z
     c_{{\rm A}\vb*{k}}
     -
     c^\dagger_{{\rm B}\vb*{k}}
     \sigma^z
     c_{{\rm B}\vb*{k}}
     \right\}
     +{\cal O}(\vb*{q})\,.
\end{align}
Therefore we have obtained the same result as
Eq.\eqref{eq:int_term_k_tot} but $\tilde{\phi}_2(0)$ is replaced by
$\tilde{\phi}_{\vb{0}}$. Finally Eq.\,\eqref{eq:H_int^A+B_discrete}
becomes
\begin{align}
  \int {\cal D}\phi
  \exp\left\{
  -\int \dd{\tau} \sum_i \left[
    \frac{2}{U}\phi_i^2
    +\vb*{c}_i^\dagger \Gamma^{5\prime} \phi_i
    \vb*{c}
    \right]+\cdots
  \right\}
  =\int {\cal D}\tilde{\phi}
  \exp\left\{
  - \int \dd{\tau} \sum_{\vb*{k}}\left[
    \frac{2N}{U} \tilde{\phi}_{\vb*{k}}\tilde{\phi}_{-\vb*{k}}
+\vb*{c}^\dagger_{\vb*{k}}
\Gamma^{5\prime}\tilde{\phi}_{\vb*{0}}\vb*{c}_{\vb*{k}} +\cdots
\right]
\right\}
\,,
\end{align}
where $\vb*{c}_i=(c_{{\rm A}i\uparrow},\,c_{{\rm
    A}i\downarrow},\,c_{{\rm B}i\uparrow},\,c_{{\rm B}i\downarrow})$.
When we make a continuum limit for space, {\it i.e.},
\begin{align}
  \frac{1}{N}\sum_i &\to \frac{1}{V}\int \dd[3]{x} \,,
  \\
  \phi_i &\to \phi(\vb*{x})\,,
\end{align}
the mass term becomes
\begin{align}
  \sum_i \frac{2}{U} \phi^2_i \to
  \int \dd[3]{x} \frac{2}{(V/N)U}\phi^2(\vb*{x})\,,
\end{align}
which agrees with Eq.\,\eqref{eq:H_A+H_B_x}.

\subsection{Mean field approximation}
\label{sec:MFA}

Another way to see the appearance of $d^5\Gamma^5$ term is use the
mean field approximation (MFA). We refer to
Refs.\,\cite{Sekine:2014xva,Chigusa:2021mci}. In the MFA,
\begin{align}
  n_{i \uparrow}n_{i\downarrow}\approx
  n_{i \uparrow}\expval{n_{i\downarrow}}
  +\expval{n_{i \uparrow}}n_{i\downarrow}
  -\expval{n_{i \uparrow}}\expval{n_{i\downarrow}}
  -\expval*{c^\dagger_{i \uparrow} c_{i\downarrow}}
  c^\dagger_{i \downarrow} c_{i \uparrow}
  -\expval*{c^\dagger_{i \downarrow} c_{i \uparrow}}
  c^\dagger_{i \uparrow} c_{i \downarrow}
  +\expval*{c^\dagger_{i \uparrow} c_{i \downarrow}}
  \expval*{c^\dagger_ {i \downarrow} c_{i \uparrow}} 
  \,.
\end{align}
The expectation values are written by introducing $\vb*{m}$\,
\begin{align}
  m^j&=\frac{1}{2}
  \sum_{\alpha,\beta}
    \expval*{c^\dagger_{i \alpha} \sigma^j_{\alpha\beta} c_{i \beta}}
  ~~~~~~~~(j=x,y)\,,
  \\
  m^z&=\frac{1}{2}
  \sum_{\alpha,\beta}
    \expval*{c^\dagger_{i \alpha} \sigma^z_{\alpha\beta} c_{i \beta}}
  +\frac{1}{2}\expval*{c^\dagger_{i \uparrow} c_{i \uparrow}
    +c^\dagger_{i \downarrow} c_{i \downarrow}}
  \,,
\end{align}
and we take
\begin{align}
  \frac{1}{2}\expval*{c^\dagger_{i \uparrow} c_{i \uparrow}
    +c^\dagger_{i \downarrow} c_{i \downarrow}} = 1\,,
  \label{eq:totalnumberdensity_MFA}
\end{align}
since the half-filling model is considered. Then, it is straightforward to
get
\begin{align}
  n_{i \uparrow}n_{i\downarrow}\approx
  \vb*{m}^2-
  \sum_{\alpha,\beta}
  c^\dagger_{i \alpha} \vb*{m}\vdot\vb*{\sigma}_{\alpha\beta} c_{i\beta}\,.
\end{align}
Extending the argument to two sublattices is trivial. Now taking
\begin{align}
    \expval*{S_{{\rm A}}^j}&=m^j=\frac{1}{2}
  \sum_{\alpha,\beta}
    \expval*{c^\dagger_{{\rm A}i \alpha} \sigma^j_{\alpha\beta} c_{{\rm A}i \beta}}
  ~~~~~~~~(j=x,y)\,,
  \\
  \expval*{S_{{\rm A}}^z}&=m^z=\frac{1}{2}
  \sum_{\alpha,\beta}
    \expval*{c^\dagger_{{\rm A}i \alpha} \sigma^z_{\alpha\beta} c_{{\rm A}i \beta}}
  +\frac{1}{2}\expval*{c^\dagger_{{\rm A}i \uparrow} c_{{\rm A}i \uparrow}
    +c^\dagger_{{\rm A}i \downarrow} c_{{\rm A}i \downarrow}}
  \,,
\end{align}
and 
\begin{align}
  \expval*{\vb*{S}_{\rm A}}+\expval*{\vb*{S}_{\rm B}}=0\,,
\end{align}
since we are interested in the AFM order, we get
\begin{align}
  U\sum_i(n_{{\rm A}i\uparrow}n_{{\rm A}i\downarrow}
  +n_{{\rm B}i\uparrow}n_{{\rm B}i\downarrow})
  \approx
  2U\sum_i\vb*{m}^2
  -U\sum_i{c}^\dagger_{{\rm A}i} \vb*{m}\vdot\vb*{\sigma}c_{{\rm A}i}
  +U\sum_i{c}^\dagger_{{\rm B}i} \vb*{m}\vdot\vb*{\sigma}c_{{\rm B}i}\,.
\end{align}
In the current model, we consider that only $m^z$ is nonzero.
Changing the variable to $\phi$ by $\phi =-Um^z$, we finally obtain
\begin{align}
  U\sum_i(n_{{\rm A}i\uparrow}n_{{\rm A}i\downarrow}
  +n_{{\rm B}i\uparrow}n_{{\rm B}i\downarrow})
  \approx
  \sum_i \frac{2}{U}\phi^2+
  \sum_i \vb*{c}_i^\dagger\phi\Gamma^{5\prime}\vb*{c}_{i}\,.
\end{align}
It is seen that both mass term for $\phi$ and $m_5\Gamma^{5\prime}$
term appear.

\section{Derivation of effective Lagrangian for $\phi$}
\label{sec:EL}

We show how to derive the effective potential~\eqref{eq:Vphi} and
stiffness $J$~\eqref{eq:J} from ${\rm Tr}(G_0\delta H)^n$ term in
Eq.\,\eqref{eq:Z_in_expand}. It corresponds to calculation of one-loop
diagram with external scalar fields those three-momenta are zero. It
is similar to the Coleman-Weinberg potential~\cite{Coleman:1973jx}, but
the the result turns out to be non-logarithmic function.  In the
following calculation, the continuum limit in the coordinate and
momentum space is taken.  We define the Fourier expansion of $G_0$ and
$\phi$ as
\begin{align}
  G_0(x-y)&=\int \frac{d^4k}{(2\pi)^4}\tilde{G}_0(k)e^{-ik\cdot (x-y)}\,, \\
  \phi(x)&=\int \frac{d^4k}{(2\pi)^4}\tilde{\phi}(k)e^{-ik\cdot x}\,,
\end{align}
where $k\cdot x=k^0x^0-\vb*{k}\cdot \vb*{x}$.

First we show how to derive the effective potential. Note that the
trace vanishes when $n$ is odd.  Thus, let us begin with $n=2$:
\begin{align}
  {\rm Tr}(G_0\delta H)^2
  &=\int \prod_{i=1}^3\dd[4]{x_i}~{\rm tr}
  \left[G_0(x_1-x_2)\phi(x_2)G_0(x_2-x_3)\phi(x_3)\right]\delta^{(4)}(x_3-x_1)
  \nonumber \\
  &=\int \frac{d^4k}{(2\pi)^4}\frac{d^4q}{(2\pi)^4}~{\rm tr}\left[
    \tilde{G}_0(q)\Gamma^5
    \tilde{G}_0(q+k)\Gamma^5
    \right]\tilde{\phi}(-k)\tilde{\phi}(k)\,.
  \label{eq:n=2term}
\end{align}
Here trace part gives
\begin{align}
  {\rm tr}\left[
    \tilde{G}_0(q)\Gamma^5
    \tilde{G}_0(q+k)\Gamma^5
    \right]
  =4\frac{\left\{q^0-\epsilon_0(\vb*{q})\right\}
    \left\{q^0+k^0-\epsilon_0(\vb*{q}+\vb*{k})\right\}
  -d_0^a(\vb*{q})d_0^a(\vb*{q}+\vb*{k})}
  {((q^0-\epsilon_0(\vb*{q}))^2-|d_0(\vb*{q})|^2)
    ((q^0+k^0-\epsilon_0(\vb*{q}+\vb*{k}))^2-|d_0(\vb*{q}+\vb*{k})|^2)}\,,
  \label{eq:n=2tr_1}
\end{align}
where index $a$ is contracted.  To give the effective potential we can
take $k=0$.  As a result, it is given in a simple form,
\begin{align}
    \left.{\rm tr}\left[
    \tilde{G}_0(q)\Gamma^5
    \tilde{G}_0(q+k)\Gamma^5
    \right]\right|_{k=0}
    =4\frac{1}
  {(q^0-\epsilon_0(\vb*{q}))^2-|d_0(\vb*{q})|^2}\,.
  \label{eq:n=2tr_2}
\end{align}
In the similar manner, higher polynomials are  calculated as
\begin{align}
     {\rm Tr}(G_0\delta H)^n
  &=\int \prod_{i=1}^{n+1}\dd[4]{x_i}{\rm tr}
   [
     G_0(x_1-x_2)\phi(x_2)
     G_0(x_2-x_3)\phi(x_3)
     \cdots
     G_0(x_n-x_{n+1})\phi(x_{n+1})
     ]\delta^{(4)}(x_{n+1}-x_1)
 \nonumber \\
  &=\int \prod_{i=1}^{n-1}\frac{d^4k_i}{(2\pi)^4}\frac{d^4q}{(2\pi)^4}~{\rm tr}
  [ 
    \tilde{G}_0(q)\Gamma^5
    \tilde{G}_0(q+k_1)\Gamma^5
    \tilde{G}_0(q+k_1+k_2)\Gamma^5\cdots
    \tilde{G}_0(q+\sum_i^{n-1}k_i)\Gamma^5
  ]
  \nonumber \\
  &~~~~~~\times \tilde{\phi}(-k_1)\tilde{\phi}(-k_2)\tilde{\phi}(-k_3)
  \cdots \tilde{\phi}(-k_{n-1})
  \tilde{\phi}(\sum_i^{n-1}k_i)
  \,,
  \label{eq:n_term}
\end{align}
where $n\ge 2$. To get the $\phi^n$ terms, we can take $k_i=0$. Then
the trace part is
\begin{align}
  \left.{\rm tr}[\cdots]\right|_{k_i=0}&=
       {\rm tr}[
         (\tilde{G}_0(q)\Gamma^5)^n
         ]
       \nonumber \\
       &=4\frac{1}{\left[
           (q^0-\epsilon_0(\vb*{q}))^2-|d_0(\vb*{q})|^2\right]^{m}}\,.
\end{align}
where $n=2m$ $(m=1,2,3,\cdots)$. Using this formula, the last term
in Eq.\,\eqref{eq:Z_in_expand} gives
\begin{align}
 -{\rm Tr}\left[\sum_{n=1}^\infty \frac{1}{n}(G_0 \delta H)^n\right]
\supset
  &-4i\int \dd[4]{x} \int \frac{d^3q}{(2\pi)^3}\frac{dq^0_E}{2\pi}
  \left[\sum_{m=1}^\infty
    \frac{1}{2m}\frac{(-1)^m\phi^{2m}}
         {\left\{(q^0_E)^2+|d_0|^2\right\}^m}\right] \nonumber \\
  &=2i\int \dd[4]{x}\int \frac{d^3q}{(2\pi)^3}\frac{dq^0_E}{2\pi}
  \log\left[1+\frac{\phi^2}{(q^0_E)^2+|d_0|^2}\right]\,.
\end{align}
Here we assume that $q^0$ integral is from $-\infty$ to $+\infty$, and
we use the Wick rotation of the the integral path as
\begin{align}
  \int_{-\infty}^{\infty}\dd{q^0}
  = i \int_{-\infty}^{\infty}\dd{q_E^0}\,,
\end{align}
Finally  the effective potential for $\phi\,(=m_5)$ is obtained as
\begin{align}
  V_\phi&=-2\int \frac{d^3q}{(2\pi)^3}\frac{dq^0_E}{2\pi}
  \log\left[1+\frac{\phi^2}{(q^0_E)^2+|d_0|^2}\right]
  +M^2\phi^2
  \nonumber \\
  &=-2\int \frac{d^3q}{(2\pi)^3}
  (\sqrt{|d_0|^2+\phi^2}-|d_0|)+M^2\phi^2\,.
\end{align}

The stiffness $J$ is obtained as an coefficient of the time-derivative
term of the dynamical field.  It is determined after the potential
minimum is found. Therefore, instead of using Eqs\,.\eqref{eq:H0} and
\eqref{eq:deltaH}, we take
\begin{align}
   H_0'(\vb*{k})&=\epsilon_0(\vb*{k}) {\bf 1}_{4\times 4}
  +\sum_{a=1}^{5}d^a(\vb*{k}) \Gamma^a\,,
  \label{eq:H0p}\\
  \delta H'(\vb*{k})&=\delta \phi \Gamma^5\,,
  \label{eq:deltaHp}
\end{align}
where now $d^5=\phi_0$ (and the others are the same) and $\delta
\phi=\phi-\phi_0$. The propagator is given similarly to
Eq.\,\eqref{eq:prop0} as
\begin{align}
  \tilde{G}(k)=
  \frac{k^0-\epsilon_0(\vb*{k})+d^a(\vb*{k})\Gamma^a}
       {(k^0-\epsilon_0(\vb*{k}))^2-|d(\vb*{k})|^2}\,,
      \label{eq:prop}
\end{align}
where $|d|^2\equiv \sum_{a=1}^4d^ad^a+\phi_0^2$. To derive the
time-derivative kinetic term, the quadratic term of $\delta H'$ should
be expanded in terms of $k^0$ (but $\vb*{k}=0$ can be taken). The
trace part is given by
\begin{align}
{\rm tr}\Bigl[
  \tilde{G}(q)\Gamma^5
  \tilde{G}(q+k)\Gamma^5
  \Bigr]\eval_{\vb*{k}=0}
&=4\frac{(q^0-\epsilon_0(\vb*{q}))^2+(q^0-\epsilon_0(\vb*{q}))k^0
  -|d_0(\vb*{q})|^2+\phi_0^2}
{((q^0-\epsilon_0(\vb*{q}))^2-|d(\vb*{q})|^2)
  ((q^0+k^0-\epsilon_0(\vb*{q}))^2-|d(\vb*{q})|^2)}\,.
 \label{eq:n=2forkin}
\end{align}
After expanding with respect to $k^0$, we get the kinetic term
\begin{align}
-{\rm Tr}\left[\frac{1}{n}(G \delta H')^n\right]\eval_{n=2}
\supset
~~&i\int\frac{d^4k}{(2\pi)^4}\delta \tilde{\phi}(-k)\delta\tilde{\phi}(k)
\int\frac{d^3q}{(2\pi)^3}
\frac{|d_0|^2}{4(|d_0|^2+\phi_0^2)^{5/2}}(k^0)^2
\nonumber \\
=&\,J\int \dd[4]{x}\delta \phi (-\partial_t^2)\delta \phi 
  \,.
\end{align}
where $\delta \tilde{\phi}$ is Fourier transformation of $\delta \phi$ and 
\begin{align}
   J=
  \int  \frac{d^3q}{(2\pi)^3}\frac{|d_0|^2}{4(|d_0|^2+\phi_0^2)^{5/2}}\,.
\end{align}

\section{$\theta$ term as chiral anomaly}
\label{sec:theta_chiral}

In a special case of the Dirac model, $\theta$ can be derived as the
path integral measure due to the chiral rotation of the electron field
$\psi$~\cite{Sekine:2014xva,Chigusa:2021mci}.  In this section we
apply the technique to the present model and clarify an issue claimed
in Ref.\,\cite{Sekine:2014xva} that $\theta$ derived in this technique
can not be applied in some circumstances.

When $B=0$ and $\epsilon_0=0$, the Hamiltonian is given by
\begin{align}
  H(\vb*{k})=Ak_x\Gamma^1+Ak_y\Gamma^2+Ak_z\Gamma^3+m\Gamma^4+\phi \Gamma^5\,.
\end{align}
In the ($\ket*{{\rm A},\sigma}$, $\ket*{{\rm B},\sigma}$)
($\sigma=\uparrow,\downarrow$) basis, the Gamma matrix is given 
as
\begin{align}
   \Gamma^{a\prime}&\equiv U_{}\Gamma^a U^\dagger \,,
  \\
  U&=\frac{1}{\sqrt{2}}\mqty({\bf 1}&~{\bf 1}~\\-{\bf 1}&~{\bf 1}~)\,.
\end{align}
To be concrete,
\begin{align}
  \Gamma^{1\prime}=\mqty(\sigma^x&0\\0&-\sigma^x)\,,~
  \Gamma^{2\prime}=\mqty(\sigma^y&0\\0&-\sigma^y)\,,~
  \Gamma^{3\prime}=\mqty(0&-i{\bf 1}\\i{\bf 1}&0)\,,~
  \Gamma^{4\prime}=\mqty(0&-{\bf 1}\\-{\bf 1}&0)\,,~
  \Gamma^{5\prime}=\mqty(\sigma^z&0\\0&-\sigma^z)\,.
\end{align}
$\Gamma^{a\prime}$ corresponds to $\tilde{\alpha}_\mu$ rotated by
$U_3$ in Ref.\,\cite{Chigusa:2021mci}, except for an extra sign in
$\tilde{\alpha}_5$. Or if one makes another transformation,
\begin{align}
  \tilde{\Gamma}^a&=\tilde{U}\Gamma^a\tilde{U}^\dagger\,,
  \\
  U&=\frac{1}{\sqrt{2}}{\rm diag}(1+i,\,1-i,\,1-i,\,1+i)\,,
\end{align}
the Dirac $\alpha$ matrices in the Dirac representation are obtained,
\begin{align}
  \tilde{\Gamma}^{1}=\mqty(0&\sigma^x\\\sigma^x&0)\,,~
  \tilde{\Gamma}^{2}=\mqty(0&\sigma^y\\\sigma^y&0)\,,~
  \tilde{\Gamma}^{3}=\mqty(0&\sigma^z\\\sigma^z&0)\,,~
  \tilde{\Gamma}^{4}=\mqty({\bf 1}&0\\0&-{\bf 1})\,,~
  \tilde{\Gamma}^{5}=\mqty(0&i{\bf 1}\\-i{\bf 1}&0)\,.
\end{align}
Here the Dirac $\alpha$ matrices corresponds to
$(\vb*{\alpha},\,\beta)=(\tilde{\Gamma}^{1},\,\tilde{\Gamma}^{2}\,
,\tilde{\Gamma}^{3}\,,\tilde{\Gamma}^{4})$. Therefore,
\begin{align}
  \tilde{U} H(\vb*{k})\tilde{U}^\dagger
  &=
  \vb*{\alpha}\vdot \vb*{k}'+\beta m+\phi \tilde{\Gamma}^5
  \nonumber \\
  &=\beta(\vb*{\gamma}\vdot \vb*{k}'+ m+\phi \gamma_5)\,,
\end{align}
where we have changed variable $\vb*{k}$ as $A\vb*{k}=\vb*{k}'$, and
introduced the Dirac gamma matrix
$\gamma^\mu=(\beta,\,\beta\vb*{\alpha})$ and
$\gamma_5=i\gamma^0\gamma^1\gamma^2\gamma^3$.  Consequently, this
Hamiltonian leads to the following action,
\begin{align}
  S=\int \dd[4]{x} \bar{\psi}
  [i\gamma^\mu (\partial_\mu -ieA_\mu) -m -i\phi\gamma_5]\psi \,,
  \label{eq:S_DiracRep}
\end{align}
where $\bar{\psi}=\psi^\dagger \gamma^0$ and we have added vector
potential $A_\mu$ of $U(1)_{\rm em}$.  Then chiral rotation of $\psi$
gives the chiral anomaly term in the
QED~\cite{Fujikawa:1979ay,Fujikawa:1980eg},
\begin{align}
  \Delta S= -\frac{\alpha}{4\pi}
  \int \dd[4]{x} \Theta F_{\mu\nu}\tilde{F}^{\mu\nu}
  =\frac{\alpha}{\pi}
\int \dd[4]{x} \Theta 
  \vb*{E}\vdot \vb*{B}
  \,,
\end{align}
where $\tilde{F}^{\mu\nu}=\frac{1}{2}\epsilon^{\mu\nu\rho\sigma}F_{\rho\sigma}$
($\epsilon^{0123}=+1$) and 
\begin{align}
  \Theta = \frac{\pi}{2}[1-{\rm sgn}(m)]{\rm sgn}(\phi)+
\tan^{-1}\frac{\phi}{m}\,.
  \label{eq:theta_chiralrotation}
\end{align}
Here we have set the domain of $\tan^{-1}$ as $[-\pi,\pi]$.  This is
consistent with Ref.\,\cite{Sekine:2014xva}, except that the sign of
$m$ and we have used single degree of freedom for electron field
$\psi$. In the model considered in
Refs.\,\cite{Sekine:2014xva,Chigusa:2021mci}, there are three Dirac
points and they expand the Hamiltonian at each Dirac points. That is
why they have three degree of freedom for electron field. (The sign of
$M_fe^{i\kappa_f\gamma_5}$ in Eq.\,(30) of Ref.\,\cite{Sekine:2014xva}
is supposed to be a plus. Comparing to the result in
Ref.\,\cite{Chigusa:2021mci} (v1 on arXiv), their result is a factor
of two smaller. In addition, it is claimed that first term is 1/2,
instead of $\pi/2$, in the topological phase.\footnote{Those are just
typos. We thank S.~Chigusa for confirming it.})  The authors of
Ref.\,\cite{Sekine:2014xva} claim that the expression of $\theta$
cannot be applied when $Um_f /\delta t'\gg 1$. ($Um_f /\delta t'$
corresponds to $\phi/m$ in our notation.)  This is misleading because
the derivation by chiral rotation in the path integral is exact and
there is no approximation. However, it should be noted that in this
calculation the cutoff of the momentum integral is taken to be
infinity. To put it more correctly, therefore, this result is reliable
when $Um_f /\delta t'$ is smaller than the cutoff momentum.  Although
the Dirac model is the low energy effective model, if the cutoff
momentum is taken to be infinity, then we get from
Eq.\,\eqref{eq:theta_phi}
\begin{align}
  \theta = {\rm sgn}(\phi)
  \Bigl[\frac{\pi}{2}-\tan^{-1}\frac{m}{|\phi|}\Bigr]\,,
\end{align}
which is the same result with Eq.\,\eqref{eq:theta_chiralrotation}.

\section{Another derivation of the Lagrangian for $\phi$}
\label{sec:Lagrangian_in_DiracRep}

In the Dirac model with $B=0$ discussed in
Appendix~\ref{sec:theta_chiral}, we can derive the Lagrangian for
$\phi$ in a way that is familiar to particle physicists. We start with
the following effective Lagrangian,
\begin{align}
  {\cal L}_{\rm eff}=
  \bar{\psi}
      [i\gamma^\mu (\partial_\mu -ieA_\mu) -m ]\psi
      -\frac{1}{2M_\phi^2}(\bar{\psi}\gamma_5\psi)^2\,.
      \label{eq:L_eff}
\end{align}
The Hubbard-Stratonovich transformation is the inverse procedure of
integrating out heavy scalar particles. In fact, above effective
Lagrangian is given by integrating out a scalar field $\phi$ that has
a Yukawa interaction with $\psi$,
\begin{align}
{\cal L}=
  \bar{\psi}
      [i\gamma^\mu (\partial_\mu -ieA_\mu) -m -i\phi \gamma_5]\psi
      -M_\phi^2\phi^2\,.
      \label{eq:L_full}
\end{align}
The last so-called four-Fermi interaction term in
Eq.\,\eqref{eq:L_eff} corresponds to Eq.\,\eqref{eq:H_int}. Taking
into account the background of the AFM order $n_{{\rm
    A}-}=-n_{{\rm B}-}$, $M^2_\phi$ is related to $U$ as
\begin{align}
  -\frac{1}{2M_\phi^2} = -\frac{UV}{4N}\,,
\end{align}
to obtain
\begin{align}
  M_\phi^2=\frac{2}{(V/N)U}\,.
\end{align}
This is the same result with Eq.\,\eqref{eq:M^2_derive}. Therefore, we
can use the Lagrangian \eqref{eq:L_full} as starting point, and
compute the effective Lagrangian for $\phi$.  As a result, we get
\begin{align}
  V_\phi^{(\rm Dirac)}=-2\int \frac{d^3q}{(2\pi)^3}
  (\sqrt{E_{\vb*{q}}^2+\phi^2}-E_{\vb*{q}})+M_\phi^2\phi^2\,,
\end{align}
where $E_{\vb*{q}}=\sqrt{\vb*{q}^2+m^2}$. This agrees with
Eq.\,\eqref{eq:Vphi} but taking $A=1$ and $B=0$. As mentioned a couple
of times, the momentum integral should have a cutoff since the Dirac
model is low energy effective model.

\end{widetext}


\end{document}